\documentclass[12pt,a4paper]{cernrep}

\usepackage{textcomp}
\usepackage[latin1]{inputenc}
\usepackage{pstricks,pst-node,pst-text}
\usepackage{amsmath}
\usepackage{graphicx}
\usepackage{axodraw}
\usepackage{cite}

\newcommand{\be}{\begin{equation}}
\newcommand{\ee}{\end{equation}}
\newcommand{\bea}{\begin{eqnarray}}
\newcommand{\eea}{\end{eqnarray}}
\newcommand{\ben}{\begin{enumerate}}
\newcommand{\een}{\end{enumerate}}

\newcommand{\crn}{\nonumber \\}

\newcommand{\ga}{\gamma}

\newcommand{\fr}{\frac}
\newcommand{\bc}{\begin{center}}
\newcommand{\ec}{\end{center}}
\newcommand{\Ga}{\Gamma}

\def\x{\chi}
\def\ti{\tilde}
\def\nt{\tilde \x^0}
\def\ch{\tilde \x^\pm}

\def\snu{\tilde\nu_\tau}
\def\stau{\tilde\tau}

\def\msnu {m_{\snu}}
\newcommand{\mnt}[1]   {m_{\tilde\x^0_{#1}}}
\newcommand{\mch}[1]   {m_{\tilde\x^\pm_{#1}}}

\newcommand{\mstau}[1] {m_{\tilde\tau_{#1}}}

\def\softsusy {{\tt SOFTSUSY\,2.0.14}}

\newcommand{\eq}[1]  {\mbox{eq.~(\ref{eq:#1})}}
\newcommand{\fig}[1] {Fig.~\ref{fig:#1}}
\newcommand{\Fig}[1] {Figure~\ref{fig:#1}}

\newcommand{\myRe} {{\Re}e}

\newcommand{\gsim}{\;\raisebox{-0.9ex}
           {$\textstyle\stackrel{\textstyle >}{\sim}$}\;}
\newcommand{\lsim}{\;\raisebox{-0.9ex}{$\textstyle\stackrel{\textstyle<}
         {\sim}$}\;}

\begin{document}

\vspace*{-18mm}
\begin{flushright}
  CERN-PH-TH/2007-197\\
  IC/2007/134\\
  LPSC 07-135
\end{flushright}
\vspace*{2mm}

\begin{center}

{\Large\bf Three-body decays of sleptons\\[2mm]
in models with non-universal Higgs masses}\\[8mm]

{\large Sabine Kraml$^{\,1}$, Dao Thi Nhung$^{\,2,3}$}\\[4mm]

{\it $^{1}$\,Laboratoire de Physique Subatomique et de Cosmologie, UJF, CNRS/IN2P3, INPG,\\
     53 Avenue des Martyrs, F-38026 Grenoble, France\\[1mm]
     $^{2}$\,Institute of Physics and Electronics, Hanoi, Vietnam\\[1mm]
     $^{3}$\,ICTP, Strada Costiera 11, I-34014 Trieste, Italy }

\vspace*{4mm}

\begin{abstract}
\noindent
We compute the three-body decays of charged sleptons and sneutrinos
into other sleptons. These decays are of particular interest in
SUSY-breaking models with non-universal Higgs mass parameters, where the
left-chiral sleptons can be lighter than the right-chiral ones, and lighter
than the lightest neutralino.
We present the formulas for the three-body decay widths together with 
a numerical analysis in the context of gaugino-mediated
SUSY breaking with a gravitino LSP.
\end{abstract}

\end{center}


\section{Introduction}

If supersymmetry (SUSY) exists at or around the TeV energy scale, experiments at 
the LHC have excellent prospects to discover it \cite{Atlas:1999fr,CMS:2007zz}.
The discovery of SUSY particles will be followed by detailed measurements 
\cite{Hinchliffe:1996iu,Branson:2001ak} of their masses and decay properties, 
with the aim to eventually determine the underlying high-scale structure of the 
theory \cite{Allanach:2004ed,Bechtle:2005vt,Kane:2006hd,Rauch:2007xp}. 
This programme will most likely require that LHC measurements be complemented 
by precision measurements at an $e^+e^-$ linear collider (ILC and/or CLIC) 
\cite{Weiglein:2004hn,Accomando:2004sz}.

Phenomenological studies and experimental simulations to assess the LHC (ILC, CLIC) 
potential are often done within minimal models of SUSY breaking, thus reducing the 
number of free parameters from more than a hundred to just a few. 
In fact, most studies are done within the framework 
of the so-called `constrained MSSM', CMSSM (often also called minimal supergravity, mSUGRA), 
which assumes that gaugino masses $m_{1/2}$, 
scalar masses $m_0$, and trilinear couplings $A_0$ are each unified at the GUT scale. 
Even if the analysis method applied is in principle model independent, the high-scale model 
leaves its imprint in, for instance, a particular mass pattern ---and thus in the resulting signatures.
It is hence important to assure that one does not miss relevant classes of signatures 
in benchmark studies. One such class of signatures, which we address in this paper, 
is three-body decays of sleptons.

In general, in SUSY-breaking models  with universal scalar and gaugino masses, 
the right-chiral charged sleptons, $\ti\ell_R$, are lighter than the 
left-chiral ones and the sneutrinos, $\ti\ell_L$ and $\ti\nu_\ell$  ($\ell = e,\mu,\tau$).  
The reason is that the renormalization group evolution of $m_{\ti\ell_R}^2$ 
is dominated by $\rm U(1)_Y$ D-term contributions, while $m_{\ti\ell_L}^2$ receives
$\rm SU(2)_L$ and $\rm U(1)_Y$ D-term corrections.
This picture changes, however, for non-universal SUSY breaking parameters at the high scale, 
especially for non-universal Higgs-mass parameters $m_{H_{1,2}}^2\not=m_0^2$, 
see e.g.~\cite{Nath:1997qm,Ellis:2002iu,Baer:2005bu}. 

Indeed for large enough  $m_{H_1}^2-m_{H_2}^2>0$, $\ti\ell_L$ and/or
$\ti\nu_\ell$ can become lighter than $\ti\ell_R$, and even lighter
than the $\nt_1$ \cite{Buchmuller:2005ma,Evans:2006sj,Buchmuller:2006nx}.
In such a setup, if R parity is conserved, the lightest SUSY particle
(LSP) should be a gravitino or axino, and the next-to-lightest one (NLSP)
a $\stau_1$ or $\ti\nu_\tau$.
As observed in \cite{Covi:2007xj}, 
SUSY cascade decays are then characterized by three-body decays of 
left-chiral sleptons at the end of the chain. 
For example, $\nt_{1,2}$ may decay into $\ti e_Le$ (or $\stau_1\tau$) 
followed by a three-body decay of the $\ti e_L$ (or $\stau_1$) into a $\ti\nu_\tau$ NLSP. 
This can considerably influence the collider phenomenology \cite{Covi:2007xj}.

In this paper, we therefore analyze the three-body decays of $\ti\ell_L$ and
$\ti\nu_\ell$ into lighter sleptons, in particular into $\stau_1$ or $\ti\nu_\tau$.
To this aim, we have computed all three-body decay widths
of sleptons (both left- and right-chiral ones) into other sleptons,
and implemented them in {\tt SDECAY}~\cite{Muhlleitner:2003vg}.
Here we give explicit formulas for the relevant decays 
of $\ti\ell_L$ and $\ti\nu_\ell$ 
and present a numerical analysis of the branching ratios
in the context of gaugino-mediated SUSY breaking.
The three-body decays of $\ti e_R$ and $\ti\mu_R$ into $\stau_1$ were
discussed in \cite{Ambrosanio:1997bq,Chou:2001nx} in the context
of gauge mediation.

The paper is organized as follows.  
In Section~2 we present the analytic expressions for the decay widths, with the generic 
structure given in Section~2.1 and the particular functions for selectrons/smuons, sneutrinos 
and staus in Sections~2.2--2.4. 
The numerical analysis is presented in Section~3, and Section~4 contains our conclusions.

\section{Three-body decay widths}

\subsection{Generic expressions}

\begin{figure}[h!]
\begin{center}
  \begin{picture}(150,130) (30,190)
    \SetWidth{0.5}
    \SetWidth{1.0}
    \Line(31,274)(90,274)
    \Line(90,274)(143,301)
    \Line(90,226)(90,274)
    \Line(142,195)(90,226)
    \Line(90,226)(142,254)
    \Text(45,280)[lb]{$P_1$}
    \Text(148,298)[lb]{$P_2$}
    \Text(148,250)[lb]{$P_3$}
    \Text(148,192)[lb]{$P_4$}
    \Text(66,248)[lb]{$X_i$}
  \end{picture}
\caption{Schematic view of a general three-body decay.}
\label{fig:threebodyscheme}
\end{center}
\end{figure}
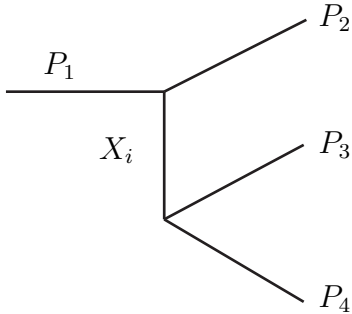

Consider a general three-body decay of electroweakly interacting 
particles $P_1 \to P_2\,P_3\,P_4$
with masses $m_{1...4}$, momenta $p_{1...4}$, and
$n_1$ exchange particles $X_i$ ($i=1,...,n_1$)
as illustrated in \fig{threebodyscheme}.

\noindent
The decay width can be written in the form
\be
 \Ga_{P_1\to P_2P_3P_4}^{}=\frac{g^4\,m_{1}}
 {32(2\pi)^3}\sum_{i,i'=1}^ {n_1}\sum_{\alpha=1}^{n_2}
 c_{ii'}^{\alpha} I_{ii'}(f^{\alpha}) \,,
 \label{eq:genericwidth}
\ee
where $g$ is the SU(2) gauge coupling and
we introduce the two-dimensional integral
\be
I_{ii'}(f^{\alpha})=\int^{ 1+r_2^2-(r_3+r_4)^2}_{ 2r_2}
 dx_{2}\int^{x_3^{\rm max}}_{x_3^{\rm min}} dx_{3}\,D_{ii'}^{X}
 f^{\alpha}(x_{2},x_{3})\,.
\ee
Here, the mass ratio $r_j:= m_j/m_1$  and integration variable $x_j:= 2p_1p_j/m_1^2$. 
The integration range $(x_3^{\rm min},x_3^{\rm max})$ is given by
\be
  x_3^{\rm min,\,max} =\frac{b\mp \sqrt{\Delta}}{2a} \,,
\ee
with
\bea
  a &=& 1- x_{2}+r_{2}^2 , \crn
  b\, &=& (2-x_{2})(1-x_{2}+r_{2}^2+r_{3}^2-r_{4}^2), \crn
  \Delta &=& (x_{2}^2-4r_{2}^2)\left[(1-x_{2}+r_{2}^2-r_{3}^2-r_{4}^2)^2
  - 4r_{3}^2r_{4}^2 \right].
\eea
The function $D_{ii'}^{X}$ is defined as
\be
D_{ii'}^{X}=\myRe\left[\left(a-r^2_{X_i}-ir_{X_i}^{}\frac{\Ga_{X_i}}{m_1}\right)^{-1}
\left(a-r^2_{X_{i'}}-ir_{X_{i'}^{}}\frac{\Ga_{X_{i'}}}{m_1}\right)^{-1}\right]\,,
\label{eq:DXii}
\ee
with $\Ga_{X_{i}}$ the decay width of the exchange particle $X_i$.
The coefficients $c^\alpha$ and functions $f^\alpha$ can easily be
obtained from the squared amplitudes of the contributing diagrams;
the sum over $\alpha=1,...,n_2$ has been introduced simply to obtain
concise expressions.
Below we give the explicit expressions for the processes relevant
to our analysis.

\subsection{Selectron and smuon decays}

\begin{figure}[t]\centering
\includegraphics[width=13cm]{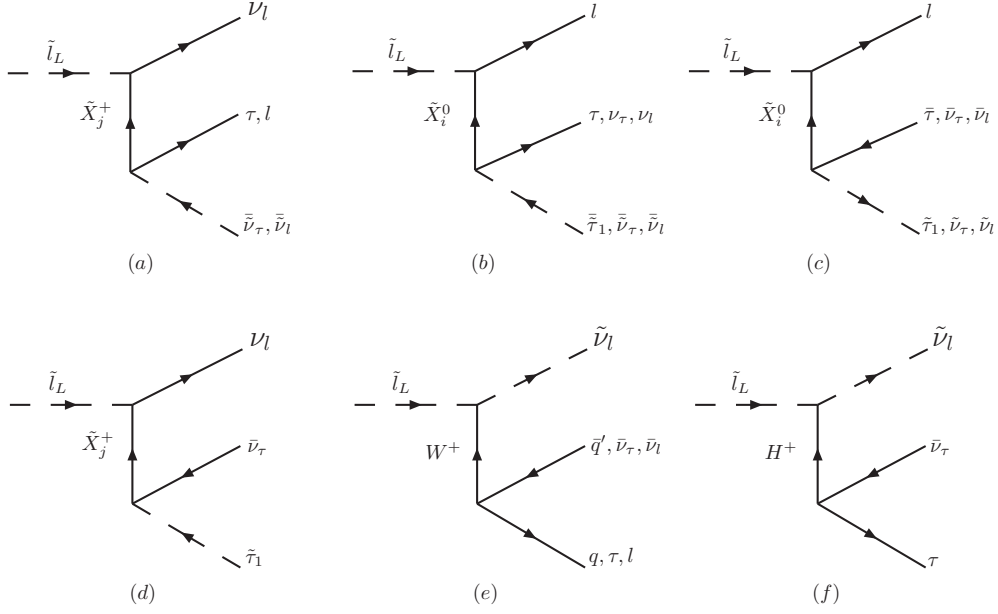}
\caption{\label{fig:diag_selectron-3body-decay}{Feynman diagrams for
the three-body decays of selectrons and smuons ($i=1,...,4$; $j=1,2$).}}
\end{figure}

Let us start with the decays of selectrons and smuons into $\ti\nu_\tau$ 
or $\stau_1$. The relevant
Feynman diagrams are shown in \fig{diag_selectron-3body-decay}.
Here and in the following we use the notation $l\equiv e,\mu$.
Moreover, $l\equiv l^-$, $\bar l\equiv l^+$, etc.
%
%
The amplitude for the decay $\tilde{l}_L\to\bar{\tilde{\nu}}_{\tau}\tau\nu_l$,
diagram (a) in \fig{diag_selectron-3body-decay}, is
\bea
 M_{\tilde{l}_L\rightarrow\bar{\tilde{\nu}}_{\tau}\tau\nu_l} &=& -ig^2\sum_{j=1}^{2}
\bar{u}(p_{\nu_l})(l_{j}^{\tilde{l}_L} P_R + k_{j}^{\tilde{l}_L}P_L)\,
 \frac{ \not{p}_{\nu_l}-\not{p}_{\tilde{l}_L}+\mch{j}}{m^2_{\tilde{l}_L}
(1-x_{\nu_l}+ r_{\nu_l}^2)-\mch{j}^2-i \mch{j}\Ga_{\ch_j}} \crn
  &&\times \, (l_{j}^{\tilde{\nu}_{\tau}} P_R +k_{j}^{\tilde{\nu}_{\tau}} P_L)
  \, v(p_{\tau})\,,
\eea
where $P_{R,L}=(1\pm\ga^5)/2$, and 
the slepton couplings to charginos are, following \cite{Belanger:2006qa},
\be
\begin{array}{ll}
  l_{j}^{\tilde{l}_L} = -U_{j1}\,,\quad &
  l_{j}^{\tilde{\nu}_{\tau}} = -V_{j1}\,, \\
  k_{j}^{\tilde{l}_L} = 0\,, &
  k_{j}^{\tilde{\nu}_{\tau}} = h_\tau U_{j2}^*\,,
\end{array} \label{eq:coupling1}
\ee
with $U$ and $V$ the chargino mixing matrices in SLHA \cite{Skands:2003cj}
notation and $h_\tau=m_\tau/(\sqrt{2}m_W\cos\beta)$
the tau Yukawa coupling.
According to \eq{genericwidth}, the decay width can be written as
($m_1=m_{\tilde{l}_L}$, $n_1=2$, $n_2=3$)
\be
  \Ga_{\tilde{l}_L\rightarrow\bar{\tilde{\nu}}_{\tau}\tau\nu_l}=
  \frac{g^4\,m_{\tilde{l}_L}}{32(2\pi)^3}
  \sum_{j,j'=1}^2\sum_{\alpha=1}^3
  c_{jj'}^{\alpha} I_{jj'}(f^{\alpha})\,,
\ee
with
\bea
c_{jj'}^1&=&l_{j}^{\tilde{\nu}_{\tau}}l_{j'}^{\tilde{\nu}_{\tau}*}l_{j}^{\tilde{l}_L}l_{j'}^
{\tilde{l}_L*} r_{\ch_j} r_{\ch_{j'}}\,,\crn
c_{jj'}^2&=& k_{j}^{\tilde{\nu}_{\tau}}k_{j'}^{\tilde{\nu}_{\tau}*}l_{j}^{\tilde{l}_L}l_{j'}^
{\tilde{l}_L*}\,,\crn
c_{jj'}^3&=&l_{j}^{\tilde{l}_L}l_{j'}^{\tilde{l}_L*}
\,\myRe\left[l_{j}^{\tilde{\nu}_{\tau}}k_{j'}^{\tilde{\nu}_{\tau}*}r_{\ch_j}
+l_{j'}^{\tilde{\nu}_{\tau}*}k_{j}^{\tilde{\nu}_{\tau}}r_{\ch_{j'}}\right], \crn
f^1&=&-1+r_{\tilde{\nu}_{\tau}}^2-r_{\tau}^2+x_{\tau}+x_{\nu_l}\,,\crn
f^2&=&1-r_{\tilde{\nu}_{\tau}}^2+r_{\tau}^2-x_{\tau}-x_{\nu_l}+x_{\tau}x_{\nu_l}\,, \crn
f^3&=&r_{\tau}x_{\nu_l}\,.
\label{eq:seL_snutau_a}
\eea
For the decay $\ti{l}_L \to \bar{\ti{\nu}}_{\tau} \nu_{\tau}l$,
\fig{diag_selectron-3body-decay}(b), we choose
$n_2=2$ and obtain 
\bea
c^1_{ii'}&=&b^{\ti{l}_L}_{i}b^{\ti{l}_L*}_{i'}a^{\ti{\nu}_{\tau}*}_{i}
         a^{\ti{\nu}_{\tau}}_{i'}\,,\crn
c^2_{ii'}&=&a^{\ti{l}_L}_{i}a^{\ti{l}_L*}_{i'}a^{\ti{\nu}_{\tau}*}_{i}
a^{\ti{\nu}_{\tau}}_{i'}r_{\nt_{i}}r_{\nt_{i'}}\,,\crn
f^1&=&1-r_{\ti{\nu}_{\tau}}^2-x_{l}-x_{\nu_{\tau}}+x_{l}x_{\nu_{\tau}}\,,\crn
f^2&=&-1+r_{\ti{\nu}_{\tau}}^2+x_{l}+x_{\nu_{\tau}}.\label{eq:sl_snutau_b}
\eea 
Here the $a$'s and $b$'s are the slepton couplings to neutralinos,
c.f.~\cite{Belanger:2006qa},
\bea
a^{\tilde{l}_L}_i &=& \frac{1}{\sqrt{2}}
   \left(\tan{\theta}_{W}N_{i1}+N_{i2}\right)\,,
\crn
b^{\tilde{l}_L}_i &=& 0\,,
\crn
a^{\stau_1}_i &=& \frac{1}{\sqrt{2}}
   \left(\tan{\theta}_{W}N_{i1}+N_{i2}\right)\cos\theta_{\stau}
   - h^*_{\tau}N_{i3}\sin\theta_{\stau}\,,
\crn
b^{\stau_1}_i &=& \sqrt{2}\tan{\theta}_WN^*_{i1}\sin\theta_{\stau}
       -h_{\tau}N^*_{i3}\cos\theta_{\stau}\,,
\eea
with $\theta_W$ the weak mixing angle, $\theta_{\stau}$ the
stau mixing angle, and $N$ the neutralino mixing matrix. 
The decay width for  
$\ti{l}_L \to \ti{\nu}_{\tau} \bar{\nu}_{\tau} l$, 
\fig{diag_selectron-3body-decay}(c), is given by 
\eq{sl_snutau_b} with the replacement 
$a^{\ti{l}_L}_{j}\leftrightarrow b^{\ti{l}_L}_{j}$.


The decays into $\stau_1$ are also mediated by neutralino and chargino exchange. 
For the process $\tilde{l}_L\rightarrow\bar{\tilde{\tau}}_1\tau l$,
\fig{diag_selectron-3body-decay}(b), we have $n_1=4$, $n_2=3$, and
\bea
c^1_{ii'}&=&a^{\tilde{l}_L}_{i}a^{\tilde{l}_L*}_{i'}
            b^{\tilde{\tau}_1}_{i}b^{\tilde{\tau}_1*}_{i'}\,, \crn
c^2_{ii'}&=&a^{\tilde{l}_L}_{i}a^{\tilde{l}_L*}_{i'}
            a^{\tilde{\tau}_1}_{i}a^{\tilde{\tau}_1*}_{i'}
            r_{\nt_i}r_{\nt_{i'}}\,,\crn
c^3_{ii'}&=&a^{\tilde{l}_L}_{i}a^{\tilde{l}_L*}_{i'} \,\myRe
\left[a^{\tilde{\tau}_1}_{i}b^{\tilde{\tau}_1*}_{i'}r_{\nt_i}
+a^{\tilde{\tau}_1*}_{i'}b^{\tilde{\tau}_1}_{i}r_{\nt_{i'}}\right], \crn
f^1&=&1-r_{\tilde{\tau}_1}^2+r_{\tau}^2-x_{\tau}-x_{l}+x_{\tau}x_{l}\,, \crn
f^2&=&-1+r_{\tilde{\tau}_1}^2-r_{\tau}^2+x_{\tau}+x_{l}\,, \crn
f^3&=&r_{\tau}x_l.
\label{eq:slLstau1}
\eea
The same expressions apply for the process 
$\tilde{l}_L\rightarrow \tilde{\tau}_1\bar{\tau} l$,
\fig{diag_selectron-3body-decay}(c), 
with $b^{\tilde{\tau}_1}_{i} \leftrightarrow a^{\tilde{\tau}_1}_{i}$ 
in $c^1_{ii'}$ and $c^2_{ii'}$.
For the decay through a virtual chargino, 
$\ti{l}_L \to \ti{\tau}_1\bar{ \nu}_{\tau} \nu_l$, 
\fig{diag_selectron-3body-decay}(d), we obtain a
simple form with $n_1=2$, $n_2=1$ and 
\bea
  c^1_{jj'}&=&l^{\ti{\tau}_1*}_{j}l^{\ti{\tau}_1}_{j'}l^{\ti{l}_L}_{j}
         l^{\ti{l}_L*}_{j'}\,,\crn
  f^1&=& 1-r_{\ti{\tau}_1}^2-x_{\nu_l}-x_{\nu_{\tau}}+x_{\nu_l}x_{\nu_{\tau}}\,. 
\eea 

We next turn to selectron and smuon decays into sneutrinos of the 
first two generations. 
The decay width for $\ti{e}_L \to \bar{\ti{\nu}}_{\mu} \mu \nu_e$ 
($\ti{\mu}_L \to \bar{\ti{\nu}}_{e} e \nu_{\mu}$), 
\fig{diag_selectron-3body-decay}(a), is in principle given by 
\eq{seL_snutau_a} with appropriate replacement of couplings. 
However, since $e$ and $\mu$ masses can be neglected, it simplifies 
to ($n_1=4$, $n_2=1$) 
\bea 
c^1_{ii'}&=&l^{\ti{l}_L}_{j}l^{\ti{l}_L*}_{j'}l^{\ti{\nu}_l}_{j}
l^{\ti{\nu}_l*}_{j'}r_{\ch_{j}}r_{\ch_{j'}}\,,\crn
f^1&=&-1+r_{\ti{\nu}_{\mu}}^2+x_{\mu}+x_{\nu_e}\,.
\eea 
For the process 
$\ti{e}_L \to \bar{\ti{\nu}}_{\mu} \nu_{\mu} e$ 
($\ti{\mu}_L \to \bar{\ti{\nu}}_{e} \nu_{e} \mu$), 
\fig{diag_selectron-3body-decay}(b), $\eq{sl_snutau_b}$ applies 
with $a^{\ti{\nu}_{\tau}}_{i} \to a^{\ti{\nu}_{l}}_{i}$. 
Likewise, the width for 
$\ti{e}_L \to \ti{\nu}_{\mu} \bar{\nu}_{\mu} e$ 
($\ti{\mu}_L \to \ti{\nu}_{e} \bar{\nu}_{e} \mu$), 
\fig{diag_selectron-3body-decay}(c), is given by $\eq{sl_snutau_b}$ 
with $a^{\ti{\nu}_{\tau}}_{i} \to a^{\ti{\nu}_{l}}_{i}$
and $a^{\ti{l}_{L}}_{i} \leftrightarrow b^{\ti{l}_{L}}_{i} $.

The decay into a sneutrino of the same flavour is more complicated 
because of interferences between different diagrams. 
For the decay $\ti{l}_L \to \bar{\ti{\nu}}_{l} l \nu_{l}$,
\fig{diag_selectron-3body-decay}(a, b) we take the number of
exchange particles $n_1=6$, with $\ti X_{1...4}=\nt_{1...4}$ and 
$\ti X_{5,6}=\ch_{1,2}$. Choosing $n_2=1$ and indices    
$i,i'=1...4$ for the neutralino exchange and $h,h'=5,6$ 
for the chargino exchange, we obtain 
\bea 
c^1_{ii'}&=& a^{\ti l_L}_{i}a^{\ti l_L*}_{i'}a^{\ti \nu_l}_{i}a^{\ti
\nu_l*}_{i'}r_{\ti X_{i}}r_{\ti X_{i'}}\,, \crn 
c^1_{hh'}&=& l^{\ti \nu_l}_{h}l^{\ti \nu_l*}_{h'}l^{\ti
l_L}_{h}l^{\ti l_L*}_{h'}r_{\ti X_{h}}r_{\ti X_{h'}}\,, \crn 
c^1_{ih}&=& c^1_{hi}=r_{\nt_i}r_{\ti X_h}\myRe \left[a^{\ti
l_L}_{i}a^{\ti \nu_l}_{i} l^{\ti \nu_l*}_h l^{\ti l_L*}_h \right]
\,, \crn 
f^1&=&-1 +r_{\ti \nu_l}^2+x_l+x_{\nu_l}\,,\label{eq:sl-sn-l-nl}
\eea 
where $l_5^{\ti l_L}=l_1^{\ti l_L}$, $l_6^{\ti l_L}=l_2^{\ti  l_L}$. 
We follow the same approach for the decay 
$\ti{l}_L \to \ti{\nu}_{l} l \bar{\nu}_{l}$,
which gets contributions from neutralino and W-boson exchange, 
see \fig{diag_selectron-3body-decay}(c,\,e);  
$n_1=5$ with $\ti X_{1...4}=\nt_{1...4}$, $\ti X_5=W$ 
and $n_2=1$ gives 
\bea c^1_{ii'}&=&a^{\ti l_L}_{i}a^{\ti l_L*}_{i'}a^{\ti
\nu_l*}_{i}a^{\ti \nu_l}_{i'}\,,  \crn 
c^1_{5i}&=&c^1_{i5}=\myRe \left[a^{\ti l_L}_{i}a^{\ti\nu_l*}_{i}\right]\,,\crn 
c^1_{55}&=&1\,,\crn
f^1&=&1-r_{\ti \nu_l}^2-x_{l}-x_{\nu_l}+x_{l}x_{\nu_l}.
\eea
W-boson exchange also leads to 
$\ti{e}_L \to \ti\nu_e \mu \bar{\nu}_{\mu}$, 
$\ti\nu_e \tau \bar{\nu}_{\tau}$ and $\ti\nu_e q \bar{q}'$, and 
analogous $\ti\mu_L$ decays. Here we have $n_1=n_2=1$. 
For $\ti{e}_L \to \ti\nu_e \mu \bar{\nu}_{\mu}$ we obtain  
\bea 
  c_{11}^1&=&1\,, \crn 
  f^1&=&1 -r_{\ti{\nu}_e}^2-x_{\mu}-x_{\nu_{\mu}}+x_{\mu}x_{\nu_{\mu}}\,, 
\eea 
which also applies for the decay into light quarks up to a colour factor 3.
Finally, for the  decay  $\ti l_L \to \ti \nu_l \tau \bar{\nu}_{\tau}$
 \fig{diag_selectron-3body-decay}(e,\,f) with W and Higgs boson exchanges,
\bea
c^{1}_{11}&=&1\,,\crn
c^2_{22}&=&2\sin^4{\beta}\cos^2{\beta}r_W^2h_{\tau}^2\,,\crn
c^2_{12}&=&\fr{1}{\sqrt{2}}\sin^2{\beta}\cos{\beta}r_{\tau}r_W^{-1}h_{\tau}(r_{\ti\nu_l}^2-1)\,,\crn
c^3_{12}&=&\fr{1}{\sqrt{2}}\sin^2{\beta}\cos{\beta}r_{\tau}r_Wh_{\tau}\,,\crn
f^1&=&\fr 1 4  r^{-4}_{W} r_{\tau}^2 \left(1 - r_{\ti \nu_l}^2 \right)^2
\left( -1 +  r_{\ti \nu_l}^2 -r_{\tau}^2+ x_{\tau} +x_{\nu_l} \right)
\crn
&&+\fr 1 2 r^{-2}_{W} r_{\tau}^2 \left(1 - r_{\ti \nu_l}^2 \right)
\left( -1 +  r_{\ti \nu_l}^2 -r_{\tau}^2+ x_{\tau} -x_{\nu_l}\right) \crn
&& +\fr 1 4 r_{\tau}^2\left(3+ r_{\ti \nu_l}^2-r_{\tau}^2+x_{\tau}-3x_{\nu_l}\right)
+1- r_{\ti \nu_l}^2-x_{\tau}-x_{\nu_l}+x_{\tau}x_{\nu_l}\,,\crn
f^2&=&-1+r_{\ti\nu_l}^2-r_{\tau}^2+x_{\tau}+x_{\nu_l}\,,\crn
f^3&=&1-r_{\ti\nu_l}^2+r_{\tau}^2+x_{\tau}-x_{\nu_l}\,,
\eea
in which we have used convention $X_1=W^+$ and $X_2=H^+$. The
remaining coefficients are zero.
Analogous expressions of course apply for $\ti l_R$ decays with the
appropriate replacements of couplings $l_{j}^{\tilde{l}_L}\to
l_{j}^{\tilde{l}_R}$, etc.

\subsection{Sneutrino decays}

\begin{figure}[t]\centering
\includegraphics[width=14cm]{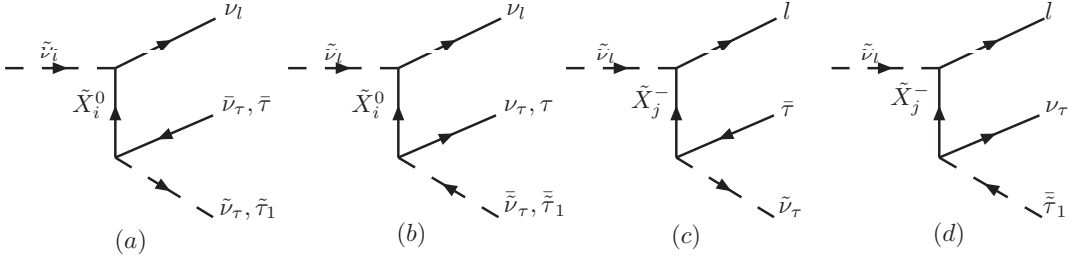}
\caption{\label{fig:diag_sneutrino}{Feynman diagrams for the three-body
decays of electron- and muon-sneutrinos ($i=1,...,4$; $j=1,2$).}}
\end{figure}

Since $\ti\nu_{e,\mu}$ are generally lighter than $\ti e_L$, $\ti\mu_L$, 
we have to consider but their decays into $\ti\nu_\tau$ or $\stau_1$.
The relevant Feynman diagrams are shown in \fig{diag_sneutrino}. 
For the invisible decays into $\ti\nu_\tau$ plus neutrinos, 
\fig{diag_sneutrino}(a,b) we find
\be
 \Ga_{\tilde{\nu}_l\to\ti\nu_{\tau}\bar{\nu}_{\tau}\nu_l,
      \bar{\ti\nu}_{\tau}\nu_{\tau}\nu_l}=
 \frac{g^4m_{\tilde{\nu}_l}}{32(2\pi)^3}\sum_{i,i'=1}^4 c_{ii'}\,I_{ii'}(f),
\ee
in terms of
\bea
  c_{ii'} &=& a^{\tilde{\nu}_l}_{i}a^{\tilde{\nu}_l*}_{i'}
              a^{\tilde{\nu}_{\tau}}_{i}a^{\tilde{\nu}_{\tau}*}_{i'}\,,\crn
  f &=& 1-r_{\tilde{\nu}_{\tau}}^2-x_{\nu_l}-x_{\nu_{\tau}}
        +x_{\nu_l}x_{\nu_{\tau}}
\eea
for $\ti\nu_l\to\ti\nu_{\tau}\bar{\nu}_{\tau}\nu_l$ and
\bea
  c_{ii'} &=& a^{\tilde{\nu}_l}_{i}a^{\tilde{\nu}_l*}_{i'}
              a^{\tilde{\nu}_{\tau}}_{i}a^{\tilde{\nu}_{\tau}*}_{i'}
              r_{\nt_i}r_{\nt_{i'}}\,,\crn
  f &=& -1+r_{\tilde{\nu}_{\tau}}^2+x_{\nu_l}+x_{\nu_{\tau}}
\eea
for $\tilde{\nu}_l\rightarrow\bar{\tilde{\nu}}_{\tau} \nu_{\tau} \nu_l$.
\noindent
For the decay $\tilde{\nu}_l\rightarrow\tilde{\nu}_{\tau} \bar{\tau} l$,
\fig{diag_sneutrino}(c), we choose $n_2=3$, leading to
\bea
c^1_{jj'}&=&l_{j}^{\tilde{\nu}_l}l_{j'}^{\tilde{\nu}_l*}l_{j}^{\tilde{\nu}_{\tau}*}l_{j'}^{\tilde{\nu}_{\tau}}\,,\crn
c^2_{jj'}&=&l_{j}^{\tilde{\nu}_l}l_{j'}^{\tilde{\nu}_l*}k_{j}^{\tilde{\nu}_{\tau}*}k_{j'}^{\tilde{\nu}_{\tau}}
r_{\ch_{j}}r_{\ch_{j'}}\,,\crn
c^3_{jj'}&=&l_{j}^{\tilde{\nu}_l}l_{j'}^{\tilde{\nu}_l*}\,\myRe\left[l_{j'}^{\tilde{\nu}_{\tau}}k_{j}^{\tilde{\nu}_{\tau}*}
r_{\ch_j}+l_{j}^{\tilde{\nu}_{\tau}*}k_{j'}^{\tilde{\nu}_{\tau}}r_{\ch_{j'}}\right]\,,\crn
f^1&=&1-r_{\tilde{\nu}_{\tau}}^2+r_{\tau}^2-x_l-x_{\tau}+x_lx_{\tau}\,,\crn
f^2&=&-1+r_{\tilde{\nu}_{\tau}}^2-r_{\tau}^2+x_l+x_{\tau}\,,\crn
f^3&=&r_{\tau}x_l \,.
\eea
For the decay $\ti \nu_l \to \ti \tau_1 \bar{\tau} \nu_l $,
\fig{diag_sneutrino}(a), we obtain with $n_2=3$:
\bea
c^1_{ii'}&=& a^{\ti \nu_l}_{i}a^{\ti \nu_l*}_{i'}a^{\ti
 \tau_1*}_{i}a^{\ti \tau_1}_{i'}\,,\crn
c^2_{ii'}&=&  a^{\ti \nu_l}_{i}a^{\ti \nu_l*}_{i'}b^{\ti
 \tau_1*}_{i}b^{\ti \tau_1}_{i'}r_{\nt_i}r_{\nt_{i'}}\,,\crn
c^3_{ii'}&=& a^{\ti \nu_l}_{i}a^{\ti \nu_l*}_{i'}\myRe \left[
a^{\ti \tau_1*}_{i}b^{\ti \tau_1}_{i'}r_{\nt_i}+a^{\ti \tau_1}_{i'}
b^{\ti \tau_1*}_{i}r_{\nt_{i'}}\right]\,,\crn
f^1&=&1-r_{\ti
  \tau_1}^2+r_{\tau}^2-x_{\nu_l}-x_{\tau}+x_{\nu_l}x_{\tau}\,,\crn
f^2&=&-1+r_{\ti \tau_1}^2-r_{\tau}^2+x_{\nu_l}+x_{\tau}\,,\crn
f^3&=&r_{\tau}x_{\nu_l}. \label{eq:sne-sL-l-ne}
\eea
The decay $\ti \nu_l \to \bar{\ti \tau}_1 \tau \nu_l,$ \fig{diag_sneutrino}(b), 
is also described by \eq{sne-sL-l-ne} with the substitution 
$a^{\ti \tau_1}_i \leftrightarrow b^{\ti\tau_1 }_i$.
Finally, for the decay $\tilde{\nu}_l\to\bar{\tilde{\tau}}_1 \nu_{\tau} l$,
\fig{diag_sneutrino}(d), we have again a simple form with $n_2=1$ and
\bea
  c_{jj'} &=& l_{j}^{\tilde{\nu}_l}l_{j'}^{\tilde{\nu}_l*}
              l_{j}^{\tilde{\tau}_1}l_{j'}^{\tilde{\tau}_1*}
              r_{\ch_{j}}r_{\ch_{j'}}\,,\crn
  f &=& -1 +r_{\tilde{\tau}_1}^2+x_l+x_{\nu_{\tau}}\,.
\eea
Last but not least we note that for $\ti\nu_\tau$ decays into $\stau_1$ 
one has to add the diagram with an off-shell $W$ boson (which is actually the 
dominant one) and the according interference terms. This is analogous to the 
$\stau_1$ decays described in the next section.

\subsection{Stau decays}

\begin{figure}[t]
\begin{center}
\includegraphics[width=12cm]{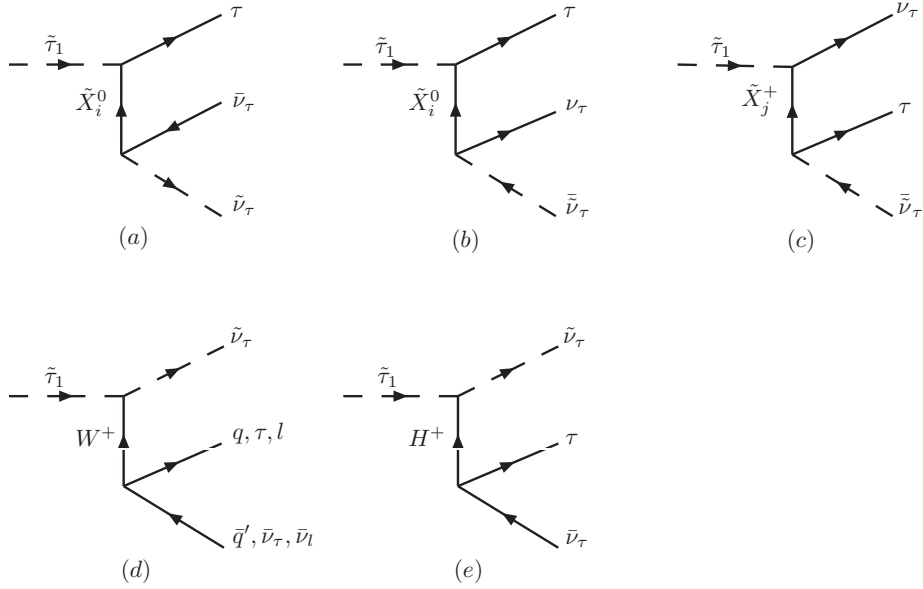}
\caption{Feynman diagrams for  $\stau_1$ decays into $\ti\nu_\tau$.
   \label{fig:diag_stau1} }
\end{center}
\end{figure}

The $\stau_1$ typically being lighter than the sleptons of the first and second 
generation, only stau decays into $\snu$ are relevant to our analysis. 
They proceed through exchange of neutralinos, charginos, $W$ or charged Higgs 
as shown in \fig{diag_stau1}.

We start with the easiest process, $\tilde{\tau}_1 \to \tilde{\nu}_{\tau}l\bar\nu_l$, 
\fig{diag_stau1}(d), for which the decay width is given by  
\be
  \Ga_{\tilde{\tau}_1 \to \tilde{\nu}_{\tau}l\bar\nu_l} = 
  \frac{g^4m_{\tilde{\tau}_1}}{32(2\pi)^3}\, I(f)
\ee
with
\be
f=\cos{\theta_{\tau}}^2\left(1-r_{\tilde{\nu}_{\tau}}^2- x_{l}
  -x_{\nu_l}+x_{l}x_{\nu_l}\right).
\ee
This result can be applied for the decay $\tilde{\tau}_1 \to \tilde{\nu}_{\tau}q\bar q'$
by multiplying with color factor 3.
For the decays involving taus in the final state, also the neutralino, chargino, 
and charged-Higgs exchange contributions have to be taken into account. 
For the process $\tilde{\tau}_1 \to \ti\nu_{\tau}\tau\bar\nu_{\tau}$,
\fig{diag_stau1}(a,\,d,\,e), we take $n_1=6$ and $n_2=6$, leading to 
\be
 \Ga_{\tilde{\tau}_1 \to \tilde{\nu}_{\tau}\tau\bar\nu_\tau}=\frac{g^4m_{\tilde{\tau}_1}}
{32(2\pi)^3}\sum_{\alpha,\alpha'=1}^6 \sum_{\beta=1}^{6}c_{\alpha\alpha'}^{\beta}
I_{\alpha\alpha'}(f^{\beta}), 
\label{eq:gamstau1A}
\ee
with the convention $X_{1...4}=\nt_{1...4}$, 
$X_{5}=W^+$ and $X_{6}=H^+$.
The coefficients and functions in \eq{gamstau1A} are 
($i,i'=1,...,4$)
\bea
c^1_{ii'}&=&a^{\tilde{\tau}_1}_{i}a^{\tilde{\tau}_1*}_{i'}
a^{\tilde{\nu}_{\tau}*}_{i}a^{\tilde{\nu}_{\tau}}_{i'}\,,\crn
c^2_{ii'}&=&b^{\tilde{\tau}_1}_{i}b^{\tilde{\tau}_1*}_{i'}
a^{\tilde{\nu}_{\tau}*}_{i}a^{\tilde{\nu}_{\tau}}_{i'}r_{X^0_i}r_{X^0_{i'}},\crn
c^2_{5i}&=&c^2_{i5}=\frac{1}{2}\cos{\theta_{\tau}}r_{W}^{-2}
r_{\tau}r_{\nt_{i}}(1-r_{\ti \nu_{\tau}}^2)
\myRe\left[b^{\tilde{\tau}_1}_{i}a^{\tilde{\nu}_{\tau}*}_{i}\right]\,,\crn
c^2_{66}&=&h_{\tau}^2a_{H\tilde{\tau}\tilde{\nu}_{\tau}}^2\sin{\beta}^2\,,\crn
c^3_{ii'}&=&r_{\tau}a^{\tilde{\nu}_{\tau}*}_{i}a^{\tilde{\nu}_{\tau}}_{i'}
\myRe\left[a^{\tilde{\tau}_1}_{i}b^{\tilde{\tau}_1*}_{i'}r_{X^0_{i'}}+
a^{\tilde{\tau}_1*}_{i'}b^{\tilde{\tau}_1}_{i}r_{X^0_i}\right]\,,\crn
c^3_{5i}&=&c^3_{i5}=\frac{1}{2}\cos{\theta_{\tau}}r_{W}^{-2}r_{\tau}^2(1-r_{\ti \nu_{\tau}}^2)
\myRe\left[a^{\tilde{\tau}_1}_{i}a^{\tilde{\nu}_{\tau}*}_{i}\right]\,,\crn
c^4_{5i}&=&c^4_{i5}=\frac{1}{2}\cos{\theta_{\tau}}\myRe
\left[a^{\tilde{\tau}_1}_{i}a^{\tilde{\nu}_{\tau}*}_{i}\right],\crn
c^{5}_{5i}&=&c^{5}_{i5}=\frac{1}{2}\cos{\theta_{\tau}}r_{\nt_{i}}r_{\tau}
\myRe\left[b^{\tilde{\tau}_1}_{i}a^{\tilde{\nu}_{\tau}*}_{i}\right],\crn
c^6_{55}&=&\cos{\theta_{\tau}}^2\,,
\eea
\bea
f^1&=&1 - r_{\tilde{\nu}_{\tau}}^2+r_{\tau}^2r_{\tilde{\nu}_{\tau}}^2
-r_{\tau}^4-x_{\tau}-x_{\bar{\nu}_{\tau}}
-x_{\bar{\nu}_{\tau}}r_{\tau}^2+x_{\tau}r_{\tau}^2+x_{\bar{\nu}_{\tau}}x_{\tau},\crn
f^2&=&-1+r_{\tilde{\nu}_{\tau}}^2-r_{\tau}^2+x_{\tau}+x_{\bar{\nu}_{\tau}},\crn
f^3&=&-1+r_{\tilde{\nu}_{\tau}}^2-r_{\tau}^2+x_{\tau},\crn
f^4&=&2-2r_{\tilde{\nu}_{\tau}}^2+r_{\tau}^2+r_{\tilde{\nu}_{\tau}}^2
r_{\tau}^2-r_{\tau}^4-2x_{\bar{\nu}_{\tau}}-2x_{\bar{\nu}_{\tau}}
r_{\tau}^2-2x_{\tau}+x_{\tau}r_{\tau}^2+2x_{\tau}x_{\bar{\nu}_{\tau}}\,,\crn
f^{5}&=&-1+r_{\tilde{\nu}_{\tau}}^2-r_{\tau}^2-x_{\bar{\nu}_{\tau}}+x_{\tau}\,,\crn
f^6&=&\fr 1 4 r^{-4}_{W}r_{\tau}^2(1-r_{\ti \nu_{\tau}}^2)^2
(-1+r_{\tilde{\nu}_{\tau}}^2-r_{\tau}^2+x_{\tau}+x_{\bar{\nu}_{\tau}})\crn
&&+\fr 1 2 r^{-2}_{W}r_{\tau}^2(1-r_{\ti \nu_{\tau}}^2)
(-1+r_{\tilde{\nu}_{\tau}}^2-r_{\tau}^2+x_{\tau}-x_{\bar{\nu}_{\tau}})\crn
&&+\fr 1 4 r_{\tau}^2(3+r^2_{\ti \nu_{\tau}}-r_{\tau}^2+x_{\tau}-3x_{\bar{\nu}_{\tau}})
+1-r^2_{\ti \nu_{\tau}}-x_{\tau}-x_{\bar{\nu}_{\tau}}+x_{\tau}x_{\bar{\nu}_{\tau}}\,,
\eea
with
\be
a_{H\tilde{\tau}\tilde{\nu}_{\tau}}=\frac{1} {\sqrt{2}}(h_{\tau}^2-1)
r_W^{-1}\sin{2\beta}\cos{\theta_{\tau}}+h_{\tau}\sin{\theta_{\tau}}
(\sin{\beta}A_{\tau}+\cos{\beta}\mu)/m_{\ti \tau_1}\,.
\ee 
The remaining coefficients are zero.
Equation (\ref{eq:gamstau1A}) also applies to the decay 
$\ti\tau_1 \to \bar{\ti\nu}_{\tau} \tau \nu_{\tau}$, \fig{diag_stau1}(b,\,c), with
$X_{1...4}=\nt_{1...4}$ and $X_{5,6}=\ch_{1,2}$:
\bea
c^1_{ii'}&=& b^{\ti \tau_1}_{i} b^{\ti
  \tau_1*}_{i'}a^{\ti\nu_\tau}_{i}a^{\ti\nu_\tau*}_{i'}\,,\crn
c^2_{ii'}&=& a^{\ti \tau_1}_{i} a^{\ti
  \tau_1*}_{i'}a^{\ti\nu_\tau}_{i}a^{\ti\nu_\tau*}_{i'}
r_{\nt_i}r_{\nt_{i'}}\,,\crn
c^2_{hh'}&=&l^{\ti \nu_{\tau}}_{h}l^{\ti \nu_{\tau}*}_{h'}
l^{\ti\tau_1}_{h}l^{\ti\tau_1*}_{h'}r_{\ti X_h}r_{\ti X_{h'}}\,,\crn
c^2_{ih}&=&c^2_{hi}= r_{\nt_i}r_{\ti X_h}\myRe
\left[a^{\ti\tau_1}_{i}a^{\ti\nu_{\tau}}_{i}l^{\ti\nu_{\tau}*}_{h}
l^{\ti\tau_1*}_{h}\right]\,,\crn
c^3_{ii'}&=&a^{\ti\nu_\tau}_{i}a^{\ti\nu_\tau*}_{i'}\myRe\left[
a^{\ti\tau_1}_{i}b^{\ti\tau_1*}_{i'}r_{\nt_i}+b^{\ti\tau_1}_{i}a^{\ti\tau_1*}_{i'}
r_{\nt_{i'}}\right]\,,\crn
c^3_{ih}&=&c^3_{hi}=r_{\ti X_h}\myRe\left[b^{\ti\tau_1}_{i}
a^{\ti\nu_{\tau}}_{i}l^{\ti\nu_{\tau}*}_{h}
l^{\ti\tau_1*}_{h}\right]\,,\crn
c^4_{hh'}&=&l^{\ti\tau_1}_{h}l^{\ti\tau_1*}_{h'}\myRe\left[l^{\ti\nu_{\tau}}_{h}
k^{\ti\nu_{\tau}*}_{h'}r_{\ti X_h}+k^{\ti\nu_{\tau}}_{h}
l^{\ti\nu_{\tau}*}_{h'}r_{\ti X_{h'}}\right]\,,\crn
c^4_{ih}&=&c^4_{hi}=r_{\nt_i}\myRe\left[a^{\ti\tau_1}_{i}
l^{\ti\tau_1*}_{h}a^{\ti\nu_{\tau}}_{i}
k^{\ti\nu_{\tau}*}_{h}\right]\,,\crn
c^5_{hh'}&=&k^{\ti\nu_{\tau}}_{h}k^{\ti\nu_{\tau}*}_{h'}l^{\ti\tau_1}_{h}l^{\ti\tau_1*}_{h'}\,,\crn
c^6_{ik}&=&c^6_{ki}=\myRe\left[b^{\ti\tau_1}_{i}a^{\ti\nu_{\tau}}_{i}k^{\ti\nu_{\tau}*}_{h}
l^{\ti\tau_1*}_{h}\right]\,,
\eea
\bea
f^1&=& 1-r_{\ti\nu_{\tau}}^2+r_{\tau}^2r_{\ti\nu_{\tau}}^2-r_{\tau}^4
-x_{\nu_{\tau}}r_{\tau}^2+x_{\tau}r_{\tau}^2-x_{\tau}-x_{\nu_{\tau}}+x_{\tau}x_{\nu_{\tau}}\,,\crn
f^2&=&-1+r_{\ti\nu_{\tau}}^2-r_{\tau}^2+x_{\tau}+x_{\nu_{\tau}}\,,\crn
f^3&=&r_{\tau}(-1+r_{\ti\nu_{\tau}}^2-r_{\tau}^2+x_{\tau})\,,\crn
f^4&=&r_{\tau}x_{\nu_{\tau}}\,,\crn
f^5&=&1-r_{\ti\nu_{\tau}}^2+r_{\tau}^2-x_{\tau}-x_{\nu_{\tau}}+x_{\tau}x_{\nu_{\tau}}\,,\crn
f^6&=&-1+r_{\ti\nu_{\tau}}^2-r_{\tau}^2+r^2_{\tau}x_{\nu_{\tau}}+x_{\tau}+x_{\nu_{\tau}}
-x_{\tau}x_{\nu_{\tau}}\,.
\eea
Here indices $i,i'=1,...,4$ and $h,h'=5,6$, and we have used the same convention 
for the couplings as in \eq{sl-sn-l-nl}. The remaining coefficients are zero.

\noindent
The three-body decays of $\snu$ into a $\stau_1$ NLSP are simply given by the inverse diagrams of 
\fig{diag_stau1}.

\section{Numerical analysis}

In order to perform a numerical analysis, we have implemented the slepton three-body decays 
in {\tt SDECAY}~\cite{Muhlleitner:2003vg}. 
The decay widths of virtual SUSY particles (charginos and neutralinos) are taken 
into account as computed by {\tt SDECAY}. The width of the charged Higgs boson is taken 
from {\tt HDECAY}~\cite{Djouadi:1997yw} through the {\tt SUSYHIT}~\cite{Djouadi:2006bz} 
interface. We have checked our code against {\tt CALCHEP}~\cite{Pukhov:2004ca} and found good agreement. 

As mentioned in the Introduction, three-body decays of left-chiral sleptons are of particular 
interest in models with non-universal Higgs-mass parameters at the high scale. 
A very attractive realization of such 
a model is the case of gaugino
mediation~\cite{Kaplan:1999ac,Chacko:1999mi}, where supersymmetry
breaking occurs on a four-dimensional brane within a
higher-dimensional theory. 
In such a setting, if gauge and Higgs superfields live in the bulk with direct couplings
the chiral superfield $S$ responsible for SUSY breaking, while squarks and sleptons 
are confined to some branes without direct coupling to $S$, we have the following 
boundary conditions at the compactification scale 
$M_C$ \cite{Chacko:1999mi}:\footnote{In our notation, 
$m^2_{H_1}=m^2_{H_d}$ and $m^2_{H_2}=m^2_{H_u}$.}  
\bea
  &&g_1 = g_2 = g_3 = g \simeq 1/\sqrt{2} \,, \crn 
  &&M_1 = M_2 = M_3 = m_{1/2} \,,\crn 
  &&m_0^2 = 0 \quad \text{for all squarks and sleptons},\crn
  &&A_0 = 0\,, \crn 
  &&\mu, B\mu, m^2_{H_{1,2}} \neq 0 \,,
  \label{eq:gauginomed}
\eea
with GUT charge normalization used for $g_1$.  The superparticle
spectrum is determined from these boundary conditions and the
renormalisation group equations.  The free parameters of the model are
hence $m_{1/2}$, $m^2_{H_1}$, $m^2_{H_2}$, $\tan\beta$, and the sign
of $\mu$; $|\mu|$ being determined by radiative electroweak symmetry
breaking.
 
The parameter ranges leading to a viable low-energy spectrum were discussed 
in \cite{Buchmuller:2005ma,Evans:2006sj} assuming $M_C =M_\text{GUT}$. 
In \cite{Buchmuller:2006nx} it was shown that either the lightest neutralino or 
the gravitino can be viable dark matter candidates in this model.  In particular,
Ref.~\cite{Buchmuller:2006nx} discussed the possibility of a gravitino
LSP with a stau or tau-sneutrino NLSP.
The collider phenomenology was discussed in \cite{Evans:2006sj} 
for the case of a neutralino LSP, and in \cite{Covi:2007xj} for the case of a gravitino 
LSP with a sneutrino NLSP. In this latter paper it was noticed that the left-chiral sleptons 
are often lighter than the $\nt_1$ and then have only three-body decays into lighter sleptons 
(the decay into the gravitino LSP being relevant only for the NLSP). 

Extending Ref.~\cite{Covi:2007xj}, we here perform a detailed numerical analysis 
of the three-body slepton decays in gaugino-mediated SUSY breaking. 
To this aim, we assume that the gravitino is the LSP and concentrate on scenarios with 
a stau or sneutrino NLSP.  Following \cite{Covi:2007xj,Buchmuller:2005ma,Buchmuller:2006nx}, 
we take $m_t=172.5$~GeV, $m_b(m_b)=4.25$~GeV and 
$\alpha_s^{\text{SM}}(M_Z)^{\overline{\text{MS}}}=0.1187$ 
as SM input parameters. Moreover, we assume $M_C = M_\text{GUT}$ and use
\softsusy\ \cite{Allanach:2001kg} to compute the sparticle and Higgs masses and 
mixing angles. In order not to vary too many parameters, we focus on 
$\tan\beta=10$ and $m_{H_2}^2=0$. 

\begin{figure}\centering
\includegraphics[width=7cm]{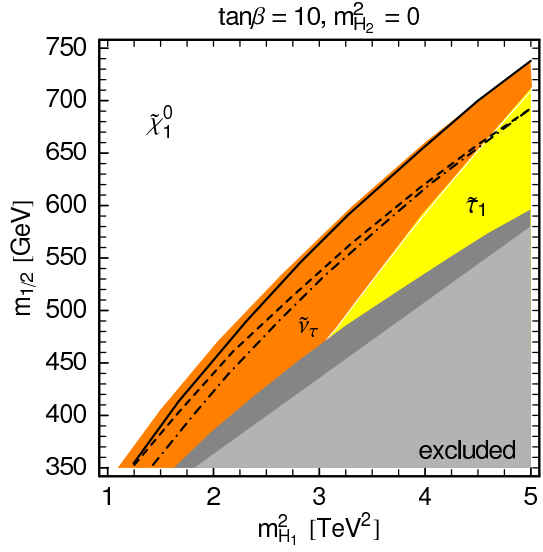}
\caption{Regions of neutralino (white), sneutrino (orange) and
stau (yellow) NLSP in gaugino mediation with a gravitino LSP.
Below the full, dashed and dash-dotted lines, respectively,
$\stau_1$, $\ti l_L$ and $\ti\nu_l$ have only three-body decays. 
The grey regions are excluded, either because no viable spectrum is obtained (light gray), 
or because $\mstau{1} < 90$~GeV (medium gray).
\label{fig:MassesContPlot}}
\end{figure}

\begin{figure}\centering
\includegraphics[width=7cm]{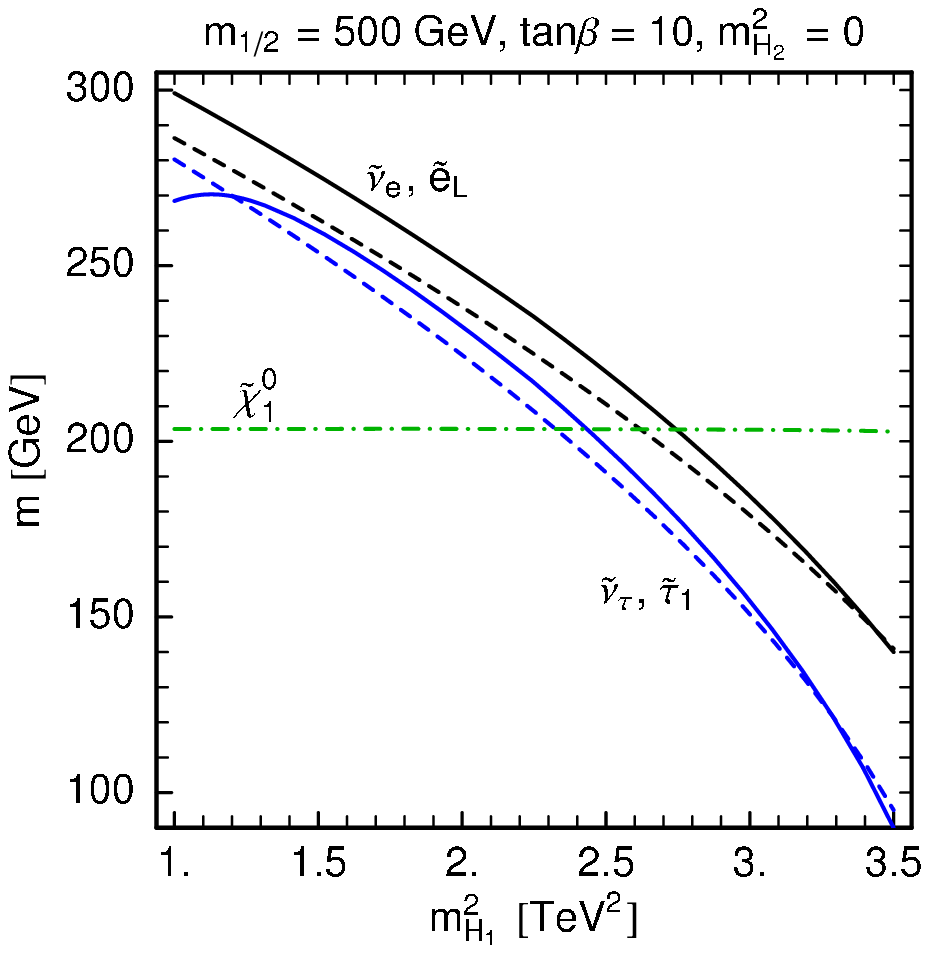}
\includegraphics[width=7cm]{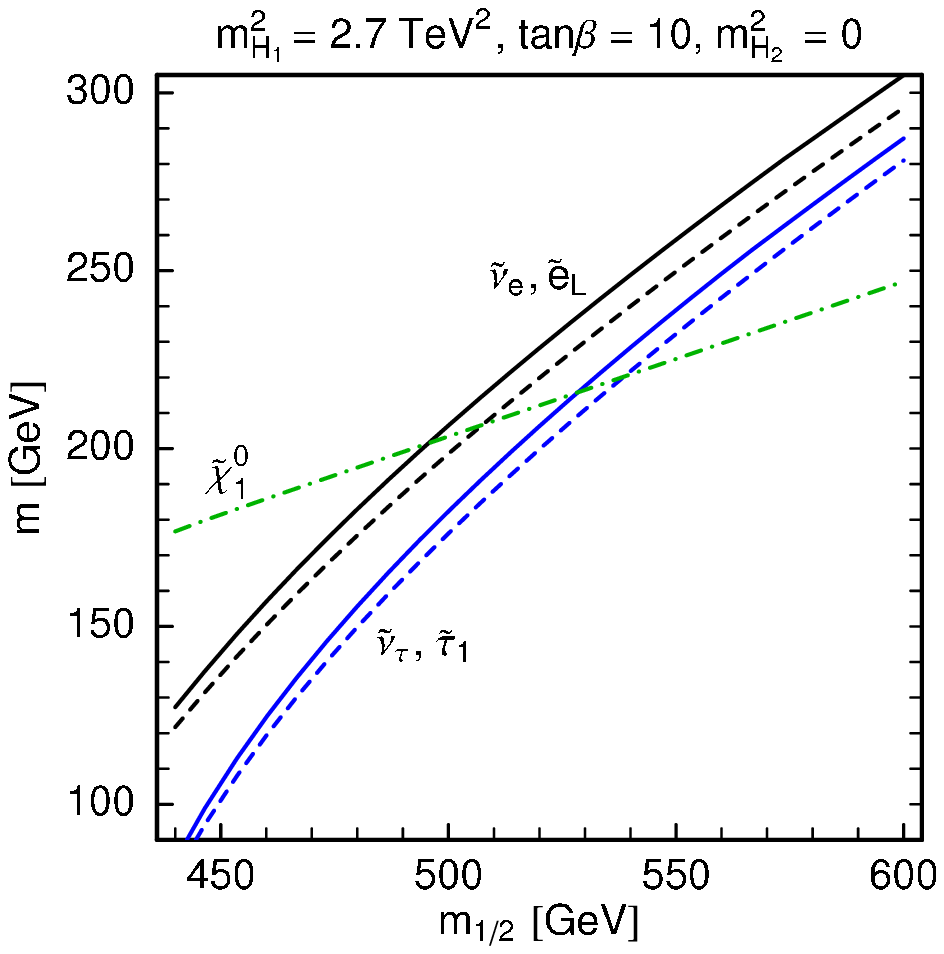}
\caption{Slepton and neutralino masses as function of $m_{H_1}^2$ (left) and $m_{1/2}$ (right). 
The dash-dotted green line is for $\nt_1$, the full and dashed blue lines are for $\stau_1$ and $\snu$, 
and the full and dashed black lines are for $\ti e_L$ and $\ti\nu_e$.
\label{fig:Masses1dim}}
\end{figure}

\Fig{MassesContPlot} shows the regions of neutralino, tau-sneutrino and stau NLSP in 
gaugino mediation with a gravitino LSP in the $m_{1/2}$ versus $m_{H_1}^2$ plane. 
The borders where three-body decays of $\stau_1$, $\ti l_L= (\ti e_L,\ti\mu_L)$, and 
$\ti\nu_l=\ti\nu_{e,\mu}$ set in, are also indicated (three-body decays of $\snu$ only occur 
in the stau NLSP region). As can be seen, three-body decays are important 
not only for $\stau_1$ and $\snu$, which always have a small mass difference~\cite{Covi:2007xj}, 
but also for $\ti e_L,\,\ti\mu_L$ and $\ti\nu_{e,\mu}$. Note also that here the 
lightest neutralino can have visible decays like, for instance, $\nt_1\to\stau_1^\pm\tau^\mp$. 
The relevant sparticle masses are shown explicitly in \fig{Masses1dim}. This figure also 
shows that the third generation sleptons are indeed always lighter than the first and 
second generation ones.


A comment is in order concerning BBN constraints. 
Gaugino mediation gives a lower bound \cite{Buchmuller:2005rt} on the gravitino mass, 
which depends on $m_{1/2}$, the number of dimensions and the compactification scale. 
For $M_C = M_\text{GUT}$ and a gluino mass of about 1 TeV, this bound is roughly 
$m_{3/2}\gsim 20$ to $0.1$~GeV for 5 to 10 dimensions \cite{Buchmuller:2005rt}. 
Hence both a $\ti\nu_\tau$ as well as a $\stau_1$ NLSP in the $\sim 100$~GeV mass range 
can be in agreement with BBN. We do not discuss this any 
further here but refer to \cite{Buchmuller:2006nx,Covi:2007xj} for details.

\begin{figure}\centering
\includegraphics[width=7cm]{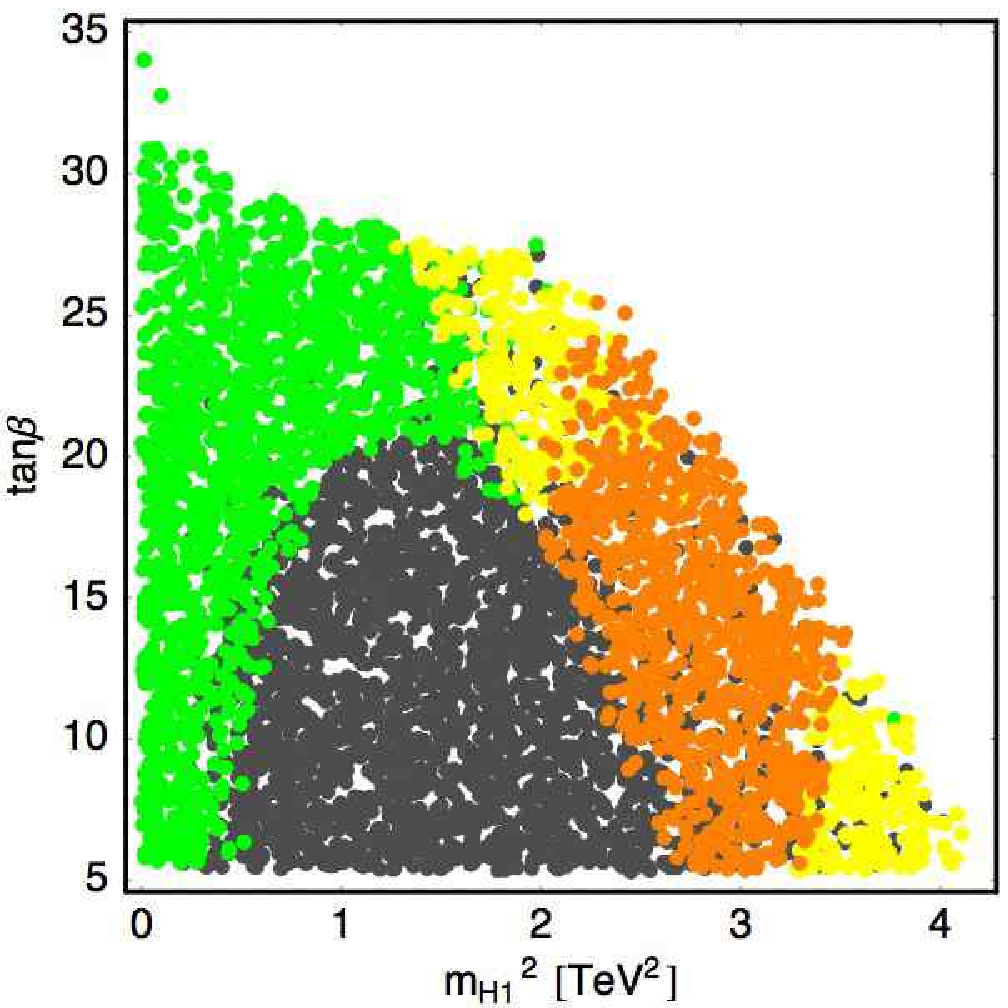}\quad
\includegraphics[width=7cm]{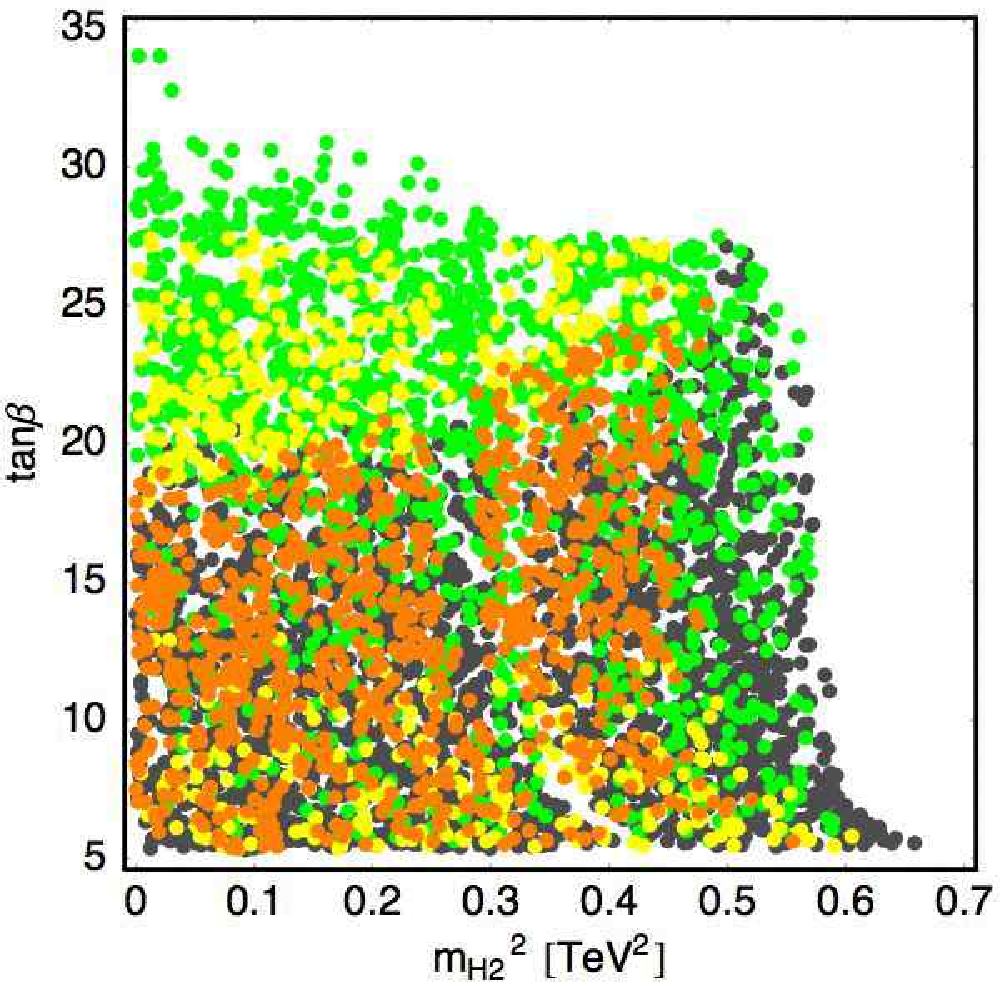}
\caption{The parameter space of gaugino mediation for $m_{1/2}=500$~GeV from a 
random scan over $m_{H_1}^2$, $m_{H_2}^2$ and $\tan\beta$, 
in the left panel marginalized over $m_{H_2}^2$ and in the right panel 
marginalized over $m_{H_1}^2$. Parameter points with a  $\nt_1$, $\stau_1$ and 
$\snu$ NLSP are shown in grey, green/yellow or orange, respectively. 
Three-body slepton decays occur for the yellow and orange points.
\label{fig:parspace}}
\end{figure}

Another comment concerns the overall paramater space of gaugino mediation. 
While we here concentrate on $\tan\beta=10$ and $m_{H_2}^2=0$, it is interesting to 
see which values these parameters could take in the general case. To this aim, 
\fig{parspace} shows results from a random scan over $m_{H_1}^2$, $m_{H_2}^2$ 
and $\tan\beta$, for $m_{1/2}=500$~GeV. The left panel shows the projection onto 
the $m_{H_1}^2$ versus $\tan\beta$ plane, and the right panel the projection onto 
the $m_{H_2}^2$ versus $\tan\beta$ plane. The grey, yellow and orange points feature 
$\nt_1$, $\stau_1$ and $\snu$ NLSPs, respectively. The green points also have 
a $\stau_1$ NLSP, however with a mass-ordering $\mstau{1}<\mnt{1}<\msnu$
for which no three-body decays of sleptons occur (in most of these cases, $\stau_1\sim\stau_R$). 
We conclude that, at $m_{1/2}=500$~GeV, three-body slepton decays occur in the range 
$m_{H_1}^2\simeq 1.3$\,--\,$4.1$~TeV$^2$, $m_{H_2}^2 \lsim 0.6$~TeV$^2$ 
and $\tan\beta\simeq 5$\,--\,27. Our choice of $\tan\beta=10$ is hence arbitrary, while the 
choice of $m_{H_2}^2=0$ is justified by the fact that the key parameter is 
$m_{H_1}^2-m_{H_2}^2$. Indeed, the yellow and orange points in \fig{parspace} 
all lie within $m_{H_1}^2-m_{H_2}^2\simeq 1.2$\,--\,$3.7$~TeV$^2$. 
 
\begin{figure}[p]\centering
\includegraphics[width=7cm]{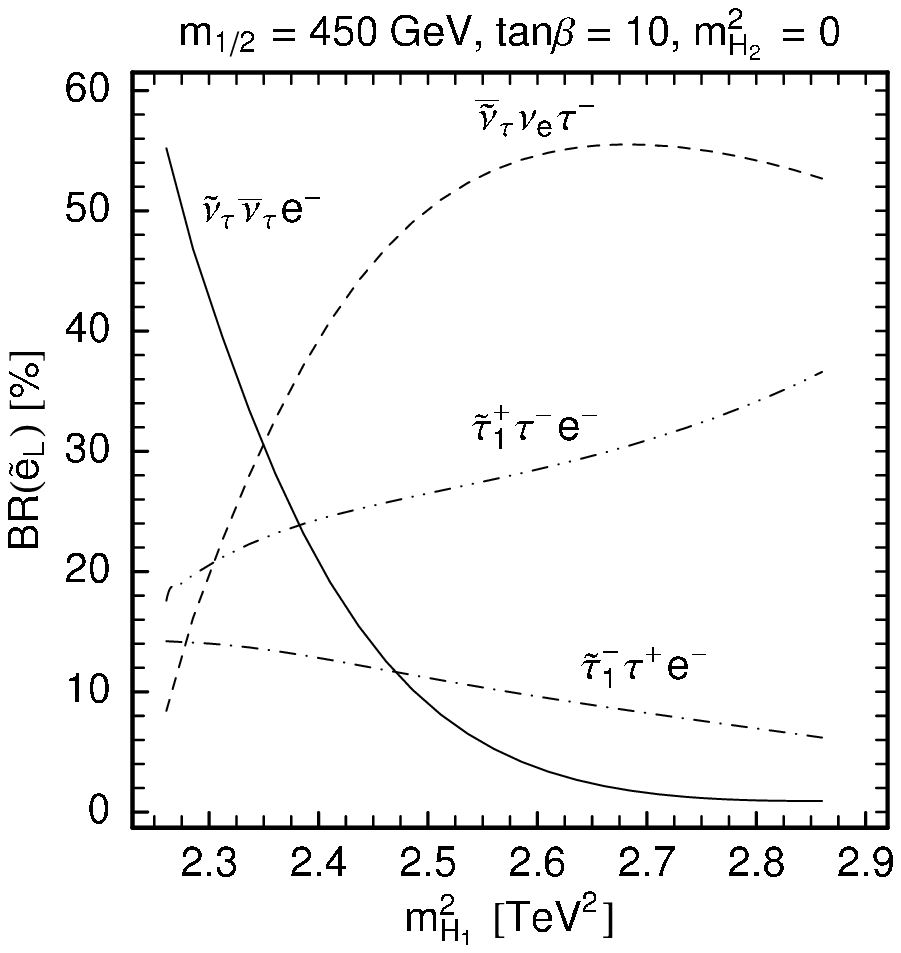}
\includegraphics[width=7cm]{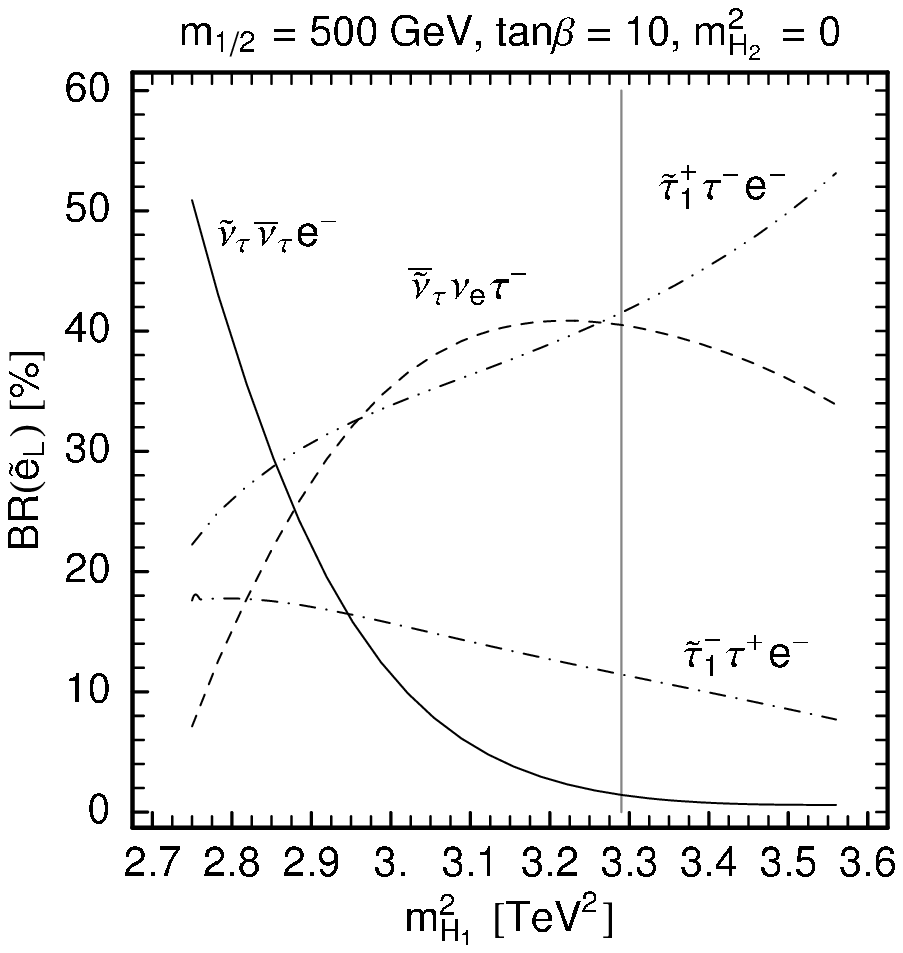}\\[4mm]
\includegraphics[width=7cm]{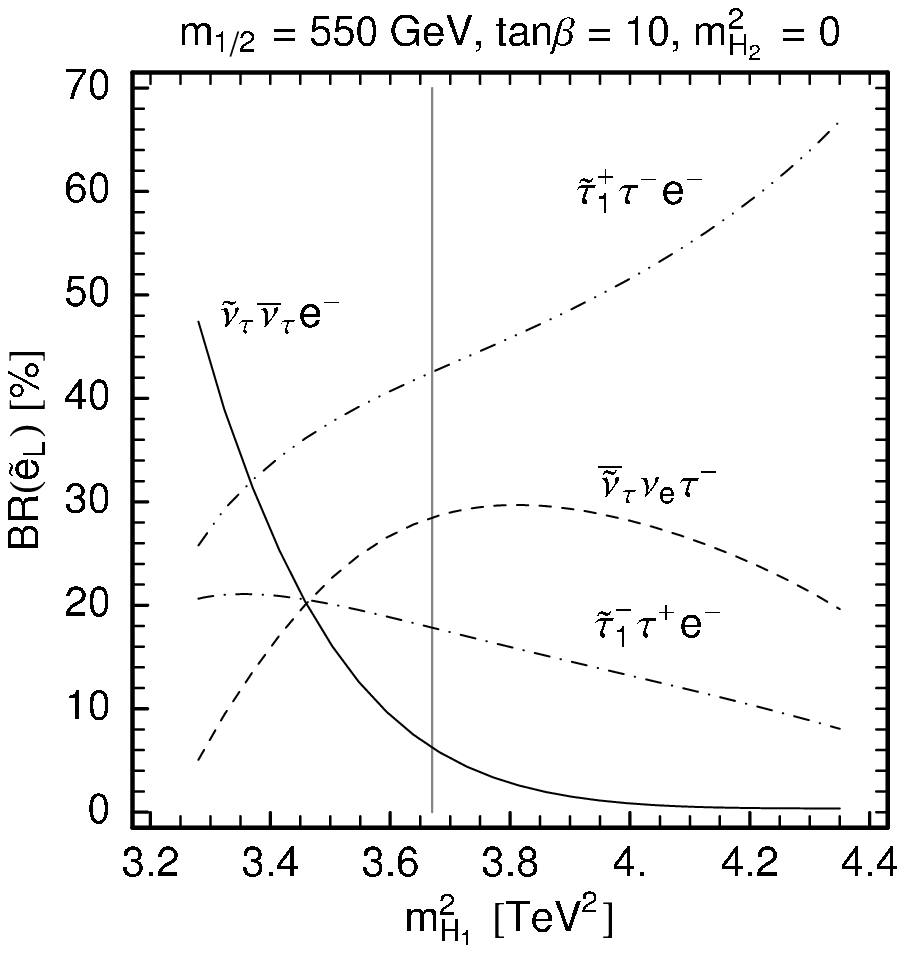}
\includegraphics[width=7cm]{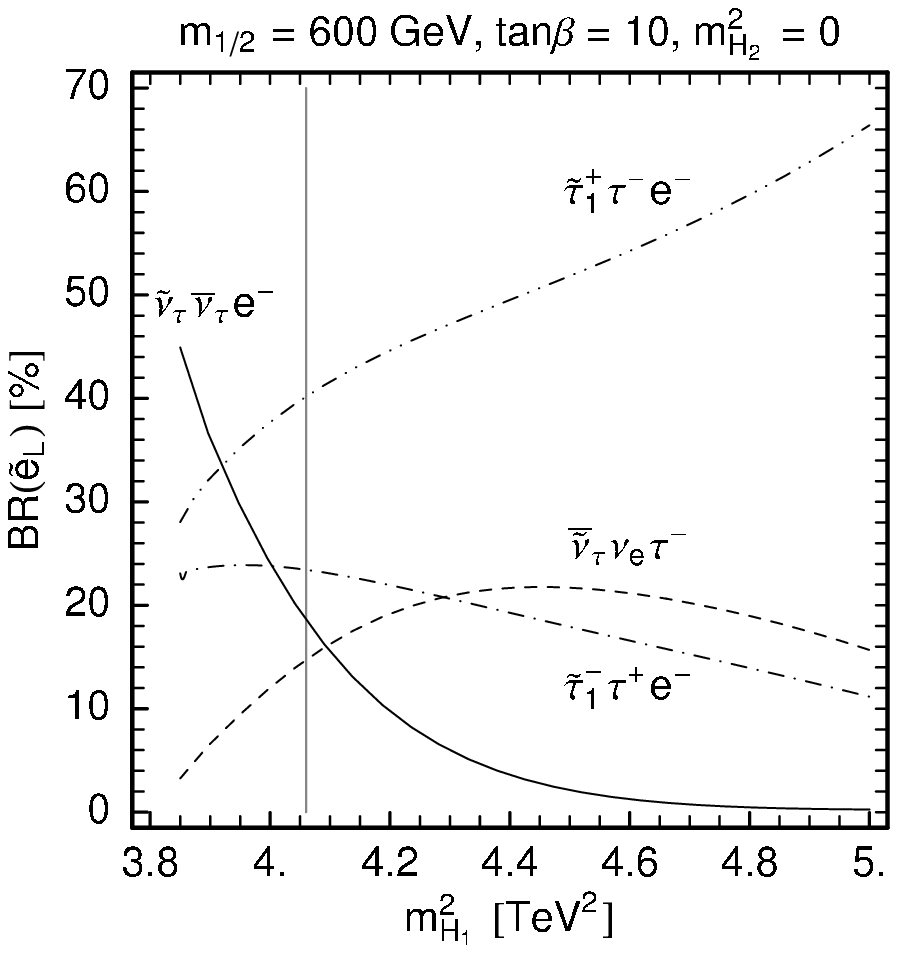}
\caption{Branching ratios of $\ti e_L$ three-body decays in \% as function of $m_{H_1}^2$, 
for $m_{H_2}^2=0$,  $\tan\beta=10$ and $m_{1/2}=450$, $500$, $550$, $600$~GeV. 
Only decay modes with sizable BRs are shown. The thin vertical line separates the $\snu$ 
NLSP region (to its left) from the $\stau_1$ NLSP region (to its right). 
In the plot for $m_{1/2}=450$~GeV, the NLSP is always the $\snu$. \label{fig:BRsell}}
\end{figure}

Let us now turn to the slepton branching ratios. The branching ratios for $\ti e_L$ decays are 
shown in \fig{BRsell} as function of $m_{H_1}^2$ for $m_{1/2}=450$, $500$, $550$ and $600$~GeV. 
The border between tau-sneutrino and stau NLSP is indicated as a thin vertical line. 
In the plot with $m_{1/2}=450$~GeV, the NLSP is always the $\snu$, and decays into it clearly dominate.
It is interesting to note, however, that with increasing $m_{H_1}^2$, i.e.\ increasing mass differences 
between $\ti e_L$ and $\snu$ as well as between $\nt_1$ and  $\ti e_L$, the decay into 
$\bar{\ti\nu}_\tau\nu_e\tau^-$ (mediated by chargino exchange) becomes more important than that into 
$\snu\bar\nu_\tau e^-$ (mediated by neutralino exchange).\footnote{The reason is that at 
   $m_{H_1}^2\simeq2.26$~TeV$^2$, where the three-body decays set in, 
   the $\nt_1$ is only slightly heavier than the $\tilde e_L$ and thus the mass ratio 
   $r_{\nt_1}=\mnt{1}/m_{\tilde e_L}\sim 1$ in \eq{DXii}, corresponding to a resonance point.
   For increasing $m_{H_1}^2$, the $\tilde e_L$ becomes lighter,  $r_{\nt_1}$ increases, 
   and the neutralino exchange becomes less important.}
If the $\ti e_L$ originates from a neutralino 
decay, $\nt_i\to\ti e_L^\pm e^\mp$, the former leads to a $\tau^\pm e^\mp E_T^{\rm miss}$ signature, which 
could mimic flavour violation. Another interesting observation is that, although the $\snu$ is lighter than the
$\stau_1$, $\ti e_L$ decays into $\stau_1$ amount to $30$--$40$\%. Here note also the asymmetry 
between the $\stau_1^+\tau^-e^-$ and $\stau_1^-\tau^+e^-$ final states. Again we have mixed-flavour 
final states. 

\begin{figure}[p]\centering
\includegraphics[width=7cm]{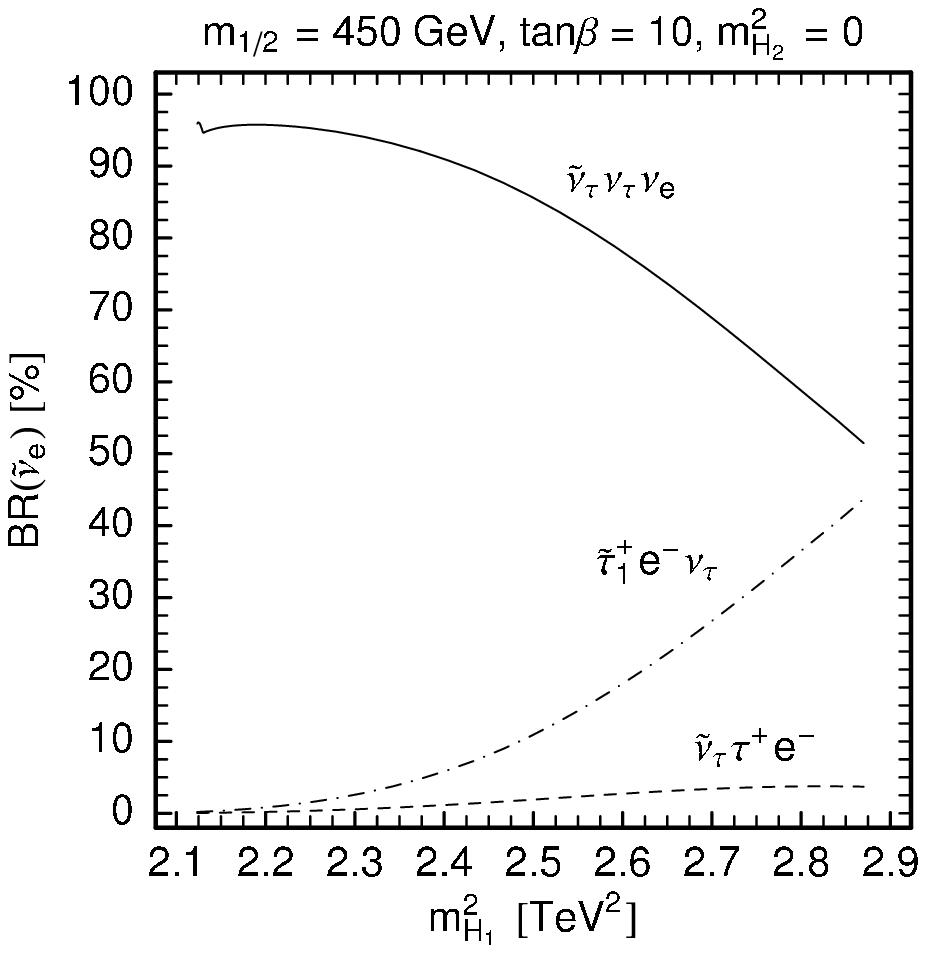}
\includegraphics[width=7cm]{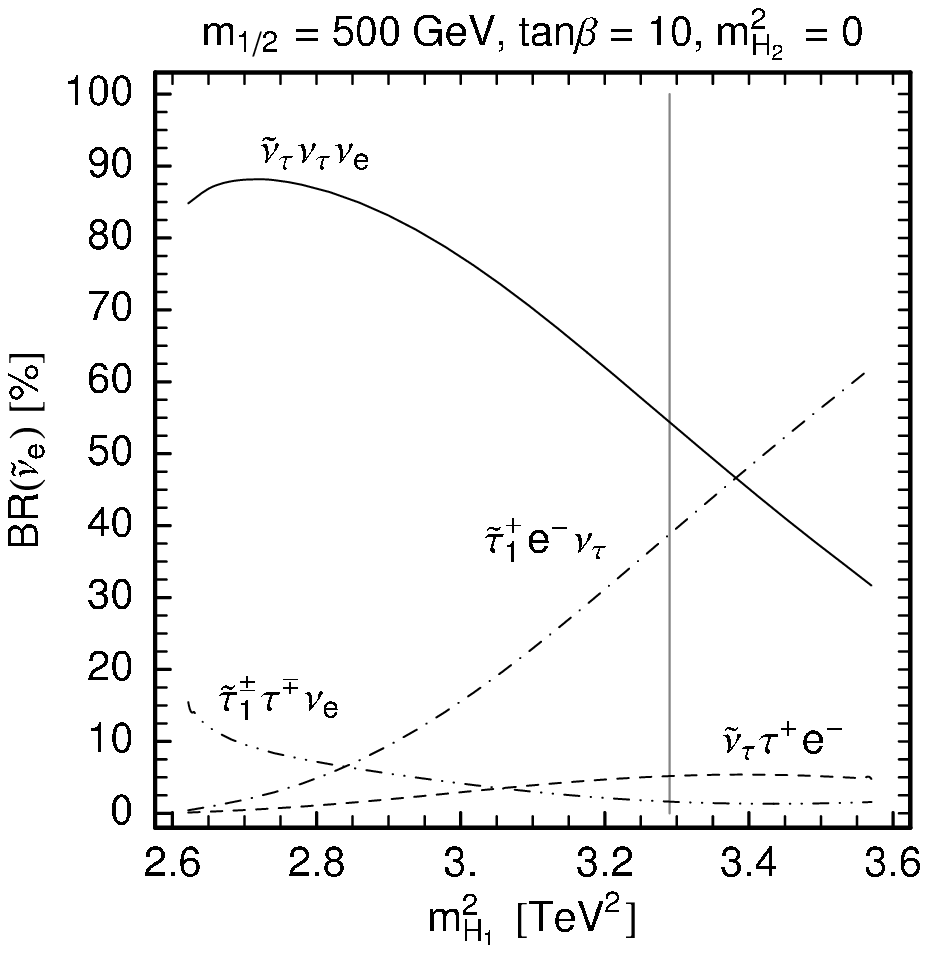}\\[4mm]
\includegraphics[width=7cm]{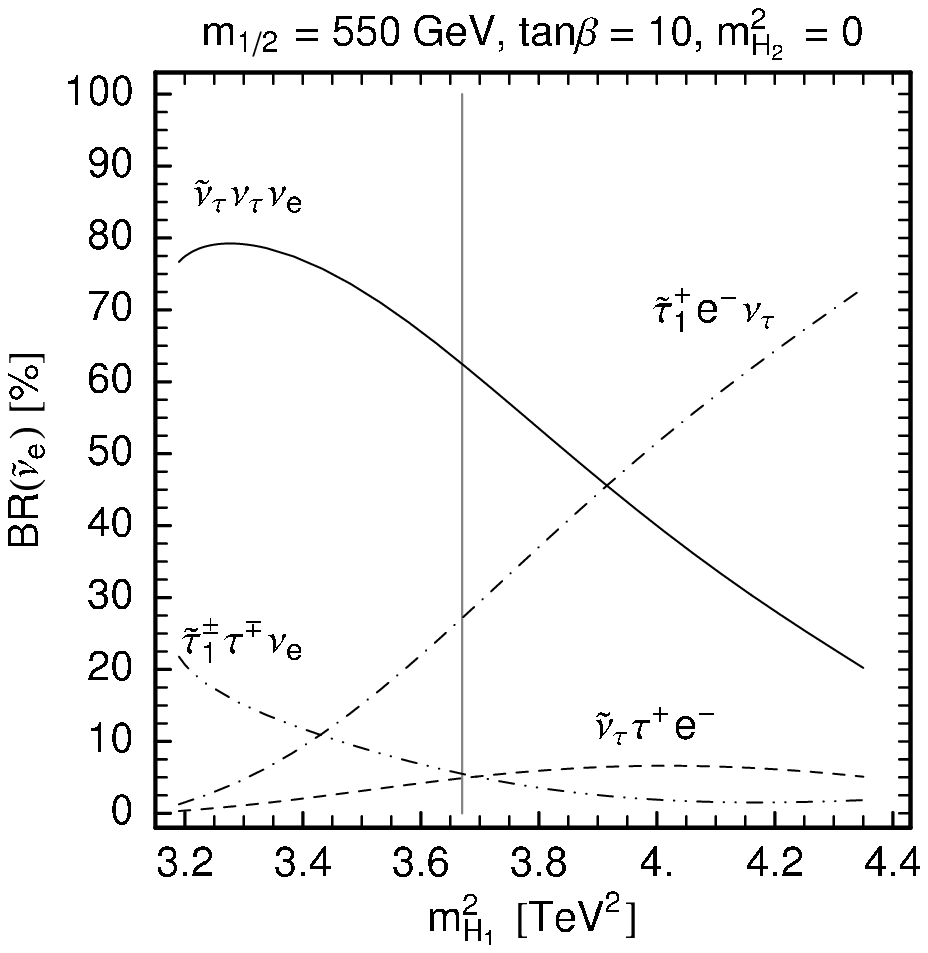}
\includegraphics[width=7cm]{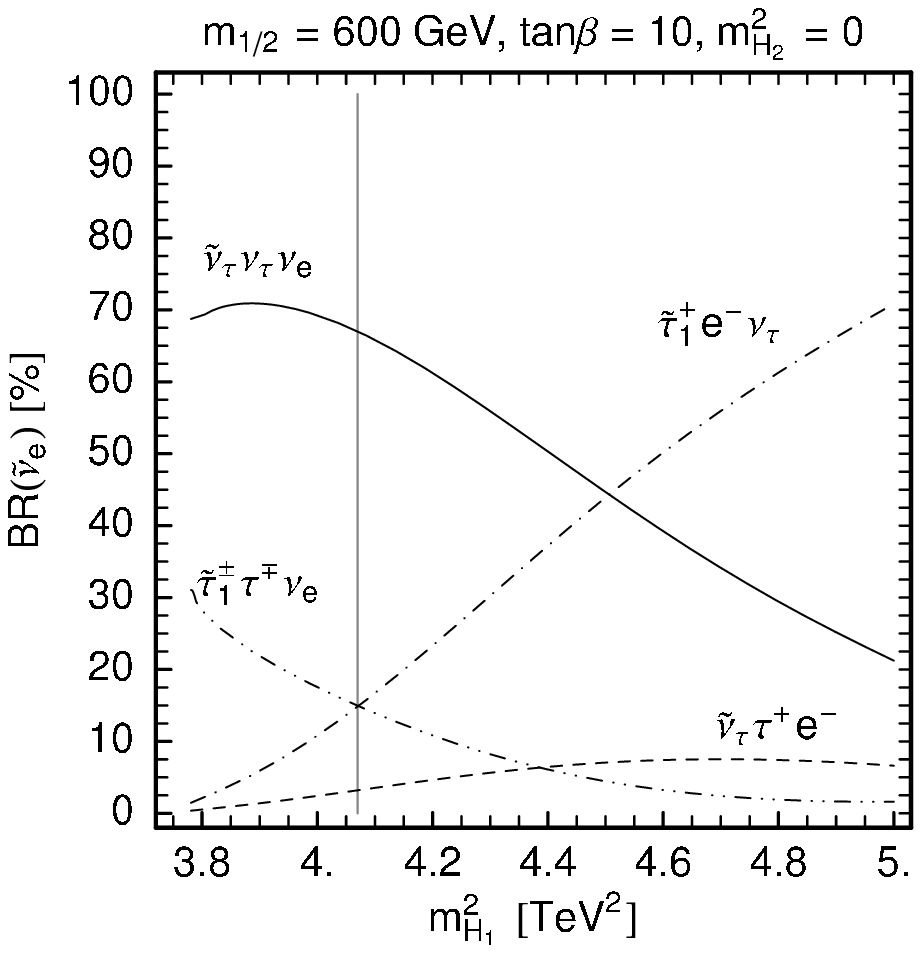}
\caption{Branching ratios of $\ti\nu_e$ three-body decays in \% as function of $m_{H_1}^2$, 
analogous to \fig{BRsell}.
\label{fig:BRsnue}}
\end{figure}

For higher values of $m_{1/2}$, there are also $\stau_1$ NLSP regions; they become larger 
with increasing $m_{1/2}$. Consequently, the $\ti e_L$ decays into $\stau_1$ get to dominate 
over those into $\snu$, c.f.\ the plots for $m_{1/2}=500$--$600$~GeV in \fig{BRsell}.
The $\ti e_L$--NLSP mass differences range from 30--50 (60)~GeV for 
$m_{1/2}=450$ ($600$)~GeV, and the total decay widths are about 
$10^{-8}$\,--\,$10^{-6}$~GeV.
Last but not least note that the decays into $\ti\nu_e$, as well as decays into $\stau_1$ 
plus neutrinos, are negligible throughout the parameter space considered here. 

We next discuss the $\ti\nu_e$ decays, for which the branching ratios are shown in 
\fig{BRsnue}. On the one hand, decays into tau-sneutrino plus neutrinos, $\ti\nu_e\to\snu\nu_\tau\nu_e$ (where the 
bars indicating anti-(s)neutrinos have been omitted for simplicity), clearly dominate for a 
$\snu$ NLSP, in which case they are invisible.
On the other hand, $\ti\nu_e$ decays into $\stau_1$ can give a visible sneutrino signature. 
The decay into $\stau_1^\pm\tau^\mp\nu_e$ can be relevant in the $\snu$ NLSP region 
for relatively large $m_{1/2}$, while the decay into $\stau_1^+e^-\nu_\tau$ becomes 
dominant in the $\stau_1$ NLSP region for large enough mass splitting. 
The total decay widths lie in the range of $10^{-7}$\,--\,$10^{-6}$~GeV.

\begin{figure}[t]\centering
\includegraphics[width=7cm]{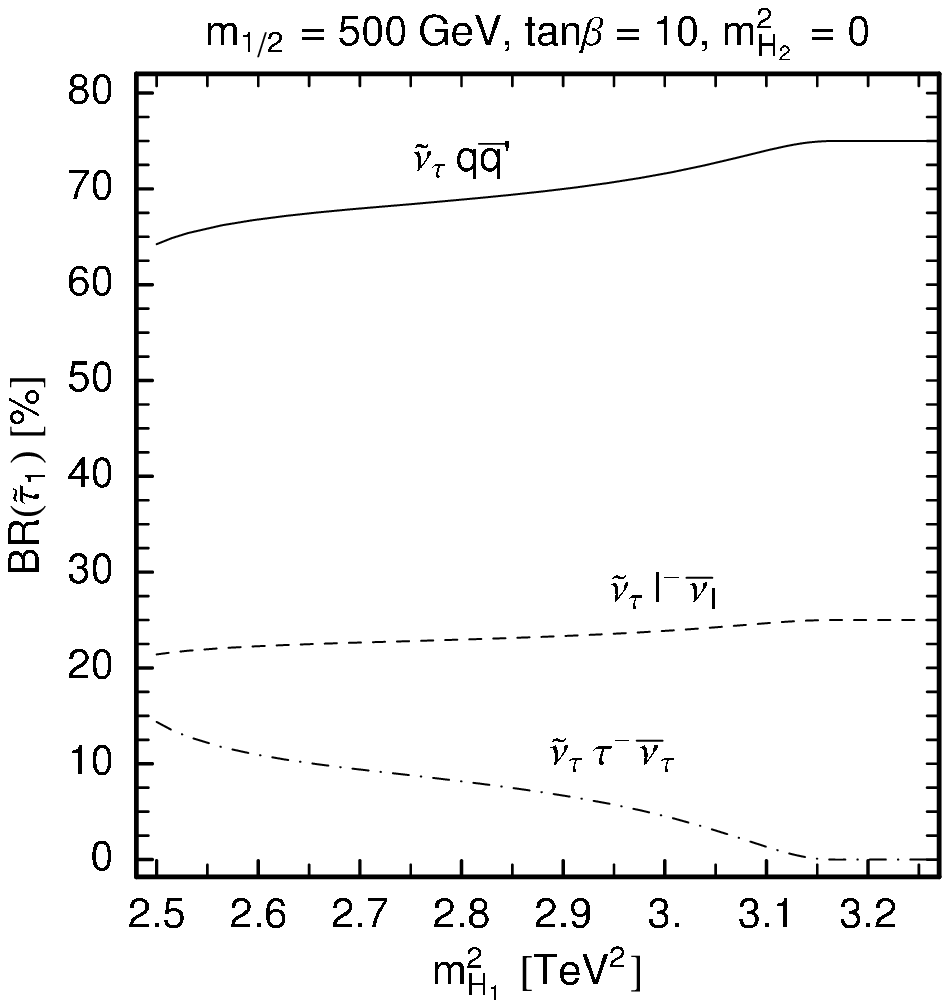}
\includegraphics[width=7cm]{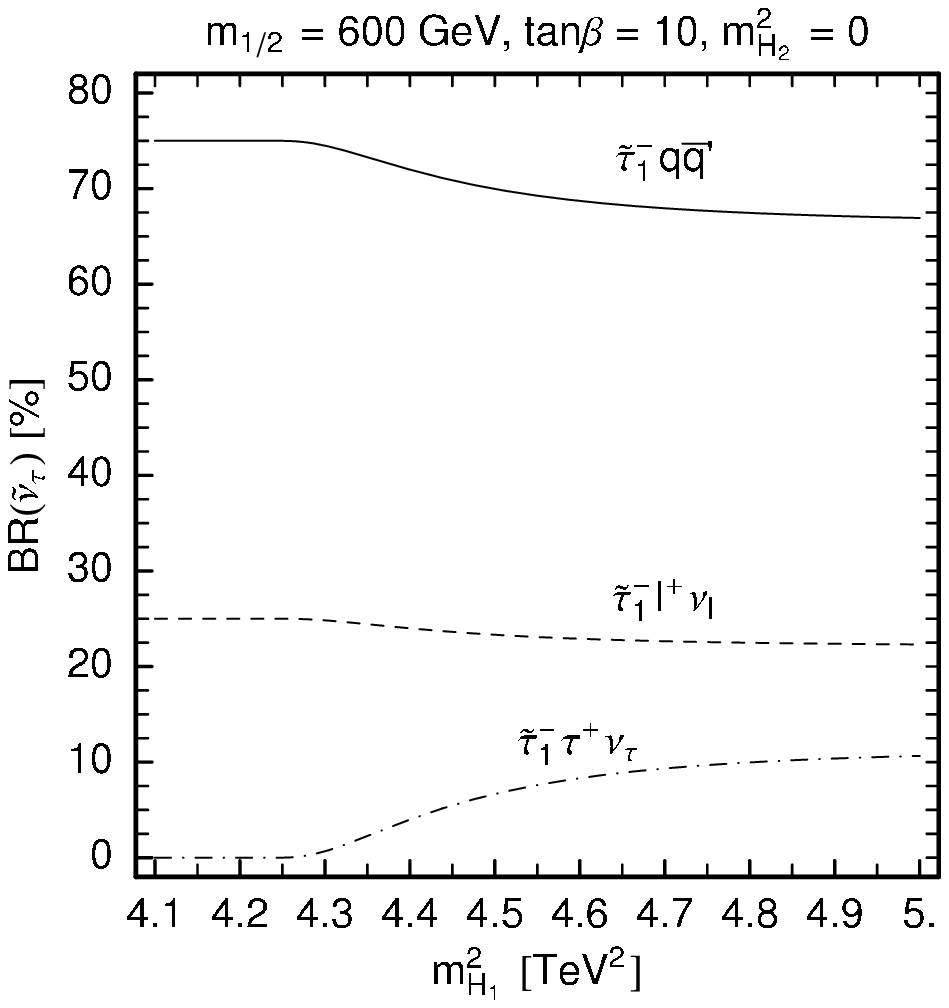}
\caption{Branching ratios in \% for $\stau_1$ decays into a $\ti\nu_\tau$ NLSP  (left) and 
for $\ti\nu_\tau$ decays into a  $\stau_1$ NLSP (right). 
\label{fig:BRstau}}
\end{figure}

The last class, $\stau_1$ decays into a $\snu$ NLSP (or $\snu$ decays into a 
$\stau_1$ NLSP) is characterized by a small $\stau_1$--$\snu$ mass difference 
of only few GeV. Here, the diagram with $W$ exchange is by far the dominant 
contribution, and the branching ratios are therefore approximately given by 
those of the $W$ boson. This is illustrated in the left panel of \fig{BRstau} for $\stau_1$ decays 
at $m_{1/2}=500$~GeV; other values of $m_{1/2}$ give very similar results.
As expected, decays into quarks have about 60--70\% branching ratio, and decays into 
electrons or muons about 20\%. 
The decay into $\snu\tau^-\bar\nu_\tau$ is, however, suppressed by the small 
$\Delta m=\mstau{1}-m_{\snu}$. One can actually see in \fig{BRstau} that it goes 
to zero for $\Delta m<m_\tau$. The total decay width 
of the $\stau_1$ is about $10^{-8}$~GeV for $\Delta m= 5$--$7$~GeV and about 
$10^{-11}$~GeV for $\Delta m\simeq m_\tau$. 
A displaced vertex only occurs for almost degenerate $\stau_1$ and $\snu$ 
($\Delta m\sim 1$~GeV or smaller). 
Analogous arguments hold for $\snu$ decays into a $\stau_1$ NLSP, as 
can be seen in the right panel of \fig{BRstau} for $m_{1/2}=600$~GeV.

\begin{figure}[t]\centering
\includegraphics[width=7.4cm]{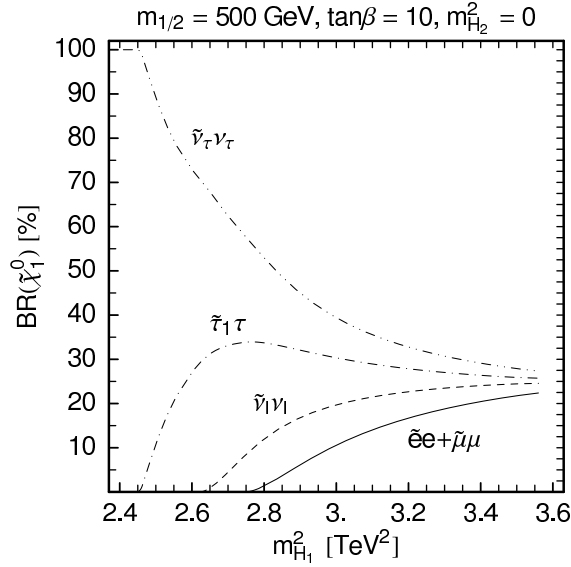}
\caption{Branching ratios of $\nt_1$ decays into sleptons (in \%) as function of $m_{H_1}^2$ 
for $m_{1/2}=500$~GeV.
\label{fig:BRnt1}}
\end{figure}

For completeness, we show in \fig{BRnt1} also the branching ratios for $\nt_1$ decays 
in the parameter range in which the three-body slepton decays are important. The 
figure is for $m_{1/2}=500$~GeV; different values of $m_{1/2}$ give similar results. 
As can be seen, decays into $\snu\nu_\tau$ dominate. Since the 
$\snu$ is the NLSP over most of the region shown, this mode is basically invisible. 
Comparing with \fig{BRsnue}, also the decays into $\ti\nu_l\nu_l$ are mostly invisible.
Visible $\nt_1$ signatures result, however, from decays into charged sleptons with 
about 30--40\% branching ratio. 


\section{Conclusions}

SUSY models with a gravitino LSP and slepton NLSP can have a quite peculiar 
collider phenomenology. This concerns not only NLSP-to-LSP decays leading to a displaced 
vertex or happening outside the detector. Also other sleptons besides the NLSP can be lighter 
than the lightest neutralino, in which case they decay through three-body modes into lighter 
sleptons, typically the NLSP. 
So far this had been considered only for right-chiral sleptons in the context of gauge mediation, 
i.e.\  three-body decays of $\ti e_R$ and $\ti\mu_R$ into a $\stau_1\sim\stau_R$ NLSP 
\cite{Ambrosanio:1997bq,Chou:2001nx}.

In this paper, we discussed an alternative possibility, namely that non-universal Higgs masses 
cause $\ti\ell_L$ and $\ti\nu_\ell$ to be lighter than $\ti\ell_R$ and $\nt_1$. To this aim we used 
a model of gaugino mediation, which has no-scale boundary conditions for the sfermion mass 
parameters and trilinear couplings, while $m_{H_{1,2}}^2\not=0$. In this model, one can naturally 
have a gravitino LSP with a mass of ${\cal O}\rm(GeV)$, and a $\stau_1\sim\stau_L$ or $\ti\nu_\tau$ NLSP.
Moreover, also $\ti e_L$, $\ti\mu_L$, and $\ti\nu_{e,\mu}$ can be lighter than the $\nt_1$. 

We hence computed all slepton three-body decays into other sleptons and implemented them in SDECAY. 
Here we presented the formulas for $\ti\ell_L$ and $\ti\nu_\ell$ decays and 
performed a numerical analysis of the branching ratios. We showed that these three-body decays 
can considerably add to the complexity of SUSY cascade decays; for instance, 
production and decays of $\ti e_L$, $\ti\mu_L$ and $\ti\nu_{e,\mu}$ typically lead to complicated 
mixed-flavour final states.  In contrast to the above-mentioned decays of right-chiral sleptons, 
here decays into sneutrinos are always important even if the NLSP is a $\stau_1$. 
The decays of $\stau_1$ into $\ti\nu_\tau$ (or vice-versa) can be very well approximated
by virtual $W$ exchange. Since the $\stau_1$--$\ti\nu_\tau$ mass difference is always small, 
the decay products tend to be soft. A dedicated experimental simulation to assess the potential 
of future colliders for such a case would be necessary. Last but not least, in the scenario 
discussed here, the $\nt_1$ has visible decays into $\ti\ell^\pm\ell^\mp$ with about 30--40\% 
branching ratio; in the case of $m_{\ti\ell_L}>\mnt{1}>m_{\ti\ell_R}$, this would be 100\%.

\section*{Acknowledgments}

We are grateful to L.~Covi and A.~Pukhov for helpful discussions.
D.T.N.\ thanks the CERN Theory Division for hospitality,
and UNESCO for financial support. S.K.\ was supported in part
by an APART grant of the Austrian Academy of Sciences.



\end{document}